\documentclass[journal, 10pt]{IEEEtran}
\IEEEoverridecommandlockouts
\usepackage{cite}
\usepackage{amsmath,amssymb,amsfonts}
\usepackage{algorithmic, algorithm}
\usepackage{graphicx}
\usepackage{textcomp}
\usepackage{xcolor}
\usepackage{amsthm}
\usepackage{multirow}
\usepackage{booktabs}
\usepackage{enumitem}
\usepackage{subcaption}
\usepackage{color}
\usepackage{bbm}
\usepackage{caption}
\renewcommand{\baselinestretch}{0.982}
\usepackage{array}

\usepackage{pifont}
\usepackage[caption=false,font=normalsize,labelfont=sf,textfont=sf]{subfig}
\usepackage{stfloats}
\usepackage{url}
\usepackage{lineno}
\usepackage{verbatim}
\usepackage{balance}
\def\BibTeX{{\rm B\kern-.05em{\sc i\kern-.025em b}\kern-.08em
    T\kern-.1667em\lower.7ex\hbox{E}\kern-.125emX}}

\renewcommand{\figurename}{Fig.}

\setlist[itemize]{noitemsep, topsep=0pt, leftmargin=*}
\setlist[enumerate]{noitemsep, topsep=0pt, leftmargin=*}

\newcommand{\blue}[1]{\textcolor{blue}{ #1}}

\begin{document}

\title{Kerckhoffs-Compliant Watermarking for Physical Design IP Protection: From Placement to Routing}

\author{Andrew B. Kahng,~\IEEEmembership{Fellow, IEEE}, and Yiting Liu,~\IEEEmembership{Member, IEEE}
\thanks{Andrew B. Kahng is with the Electrical and Computer Engineering Department and the Computer Science and Engineering Department, University of California at San Diego, La Jolla, CA 92093 USA (e-mail: abk@ucsd.edu).}
\thanks{Yiting Liu is with the Computer Science and Engineering Department, University of California at San Diego, La Jolla, CA 92093 USA (e-mail: yil375@ucsd.edu).}
}

\maketitle

\begin{abstract}
Physical design (PD) intellectual property (IP) is a valuable artifact 
of modern VLSI implementation. It includes optimized cell placement, 
clock distribution, and routing decisions produced by carefully 
tuned PD flows. As access to PD tools expands, 
unauthorized reuse of placed-and-routed databases becomes 
an increasing concern.
Existing PD watermarking methods
either protect only one PD stage or rely on hidden construction details,
leaving them vulnerable to a white-box adversary.
In this work, we develop \emph{PDMarks}, a Kerckhoffs-compliant watermarking 
framework whose security depends only on a secret key. 
\emph{PDMarks} embeds ownership evidence across multiple stages of the PD flow, 
including placement, clock tree synthesis (CTS), and routing. 
All watermark instances and target values are deterministically 
derived from a 32-byte secret key using HMAC-SHA256, 
enabling consistent embedding and verification.
\emph{PDMarks} has been integrated into
OpenROAD-flow-scripts. 
Experiments on NanGate45 and ASAP7 designs show that
\emph{PDMarks} outperforms prior physical design watermarking methods
by providing much stronger ownership evidence with comparable
or smaller PPA overhead. 
The approximate joint all-stage coincidence probability is below
$10^{-32}$ for every evaluated design. Wrong-key 
and attack evaluations further show that incorrect 
keys do not reproduce the complete ownership proof 
and that weakening the watermark requires broad 
perturbation of the protected implementation.

\end{abstract}


\section{Introduction}
\label{sec:introduction}

Physical design (PD) transforms a synthesized netlist into a
placed, clocked, and routed layout database from which a chip
can be manufactured. Beyond realizing circuit functionality,
this database records physical implementation effort: flow
configurations, constraint tuning, timing-closure iterations,
and power-performance-area (PPA) tradeoffs. 
Hence, the resulting physical implementation constitutes a high-value
design artifact.
With the increasing accessibility of PD flows, particularly 
through open-source infrastructures such as 
OpenROAD-flow-scripts~\cite{ORFS}, the risk of 
unauthorized reuse of placed-and-routed design databases has grown significantly, 
making the protection of PD-level IP an urgent concern.

The setting of PD-level IP protection differs fundamentally from classical design IP 
protection~\cite{Oliveira01}~\cite{HongP98}, 
where the protected object is typically the functionality, RTL, 
or gate-level netlist. 
At the PD level, the IP lies in the quality of implementation achieved 
through iterative physical-design optimization. 
Two layouts derived from the same netlist may exhibit substantial 
differences in timing, power consumption, and manufacturability. 
These implementation choices and their resulting 
quality are not recoverable from upstream representations alone. 
Therefore, an ownership mechanism for PD IP must bind directly to 
the physical implementation and remain detectable after standard post-layout 
optimizations.

{\em Watermarking} provides a practical approach for establishing post hoc ownership, i.e., as creator,
of such artifacts. The design owner 
embeds a signature during the PD flow.
The signature can later be extracted from a suspect layout 
and verified using a secret key.
For physical design IP protection, a natural and rigorous 
security requirement is \emph{Kerckhoffs's principle}~\cite{Kerckhoffs83}: 
the security of the system should depend only on the secrecy of the key, 
not on the secrecy of the algorithm. Accordingly, we adopt a 
white-box threat model in which the adversary has full knowledge 
of the watermarking algorithms, tool, source code, 
and design flow. Only the owner's secret key remains unknown. 
This assumption is particularly relevant for open-source flows, 
where tool implementations, command sequences, 
and intermediate checkpoints are fully accessible.

Existing physical design watermarking methods do not fully
satisfy the Kerckhoffs requirement. Constraint-based methods insert
additional constraints before optimization is solved~\cite{KahngMMP98}~\cite{KahngLMM01}. 
Post-layout modification methods create structural
patterns after layout generation~\cite{SunGX06}~\cite{CaiGBX07},
which a Kerckhoffs adversary can search for and remove. More recent methods improve
fidelity by choosing low-impact placement regions:
ICMarks~\cite{ZhangRPK25} uses a scoring function, while
AutoMarks~\cite{ZhangRPK24}~\cite{ZhangRPK25TODAES} uses a 
graph neural network (GNN) to
accelerate region search. These approaches are robust against
several placement-level attacks, but the watermark remains
localized in placement and depends on private construction
artifacts such as selected regions, signatures, parameters, or
intermediate placements. They leave clock tree 
synthesis (CTS) and routing outside
the ownership proof. This single-stage focus creates a
practical weakness. Modern PD flows involve multiple stages, 
each resolving different degrees of freedom in the 
physical implementation.
An adversary capable of re-running or locally repairing a 
specific stage can selectively disrupt embedded marks 
while preserving the majority of the stolen implementation. Consequently, a robust
watermarking strategy should distribute ownership evidence
across multiple stages and make the watermark objects
indistinguishable from normal design objects.

In this work, we propose \emph{PDMarks}, a Kerckhoffs-compliant watermarking 
framework for physical design IP protection that spans placement, 
CTS, and routing. \emph{PDMarks} embeds ownership evidence using intrinsic 
degrees of freedom at each stage: (i) the left-to-right ordering 
of selected same-row cell tuples after detailed placement, 
(ii) the sequential-fanout parity of selected local clock 
buffers during CTS, and (iii) the wrong-way routing behavior of 
selected signal nets during routing. All watermark instances are 
deterministically derived from a single 32-byte secret key using 
HMAC-SHA256, enabling consistent embedding and verification while 
binding all available stages into a unified ownership proof.

The main contributions of this paper are as follows.
\begin{itemize}
\item To the best of our knowledge, this is the first Kerckhoffs-compliant 
watermarking framework for physical design that jointly protects 
placement, CTS, and routing stages. 
Its security depends on
the owner's key even when the watermarking algorithms, source code,
tool integration, and physical design flow are fully known.
\item We develop three stage-specific watermarking mechanisms 
that are structurally entangled with the physical design solution. 
They leverage natural physical-design degrees of freedom and are
difficult to remove without inducing significant perturbations to
timing, power, or routability.
\item We introduce a unified secret-key architecture based on
domain-separated HMAC-SHA256 evaluations. A single key derives
independent stage keys and stage-level watermark instances while
binding all three stages into a single ownership proof.
\item We implement \emph{PDMarks} within the OpenROAD 
infrastructure~\cite{AjayiCFHH2019}, integrating watermark 
embedding and verification directly into the PD flow 
rather than applying post hoc modifications.
\end{itemize}

The remaining sections are organized as follows.
Section~\ref{sec:preliminaries} introduces
background, problem formulation, and the attack model.
Section~\ref{sec:related_work} reviews prior physical design
watermarking. Section~\ref{sec:methodology} presents the
three watermarking mechanisms and ownership verification procedure.
Section~\ref{sec:experiments} reports experimental results.
Section~\ref{sec:attack_evaluation} evaluates attack resilience.
We conclude the paper in Section~\ref{sec:conclusion}.

\section{Preliminaries}
\label{sec:preliminaries}

This section briefly introduces the physical design flow,
states the watermarking requirements, 
problem formulation, and attack models. 
Table~\ref{tab:notation} summarizes the notation used 
in this paper.

\begin{table}[htp]
\caption{Notation and Terminology.}
\label{tab:notation}
\centering
\begin{tabular}{lp{0.70\columnwidth}}
\hline
Symbol & Definition \\
\hline
$K$, $\hat{K}$    & 32-byte secret key, and claimed key, used during verification \\
$K_s$              &  Stage key at stage $s$\\
$P, C, R$         & Placement, CTS, Routing \\
$s$               & Watermarking stage, $s \in \{P, C, R\}$ \\
$\mathcal{A}$  & Set of available watermarking stages \\
$\mathcal{D}_0$   & Input design state before physical implementation \\
$d_m$               & Degradation for quality metric $m$ \\
$\mathcal{F}$     & Physical design flow \\
$\mathcal{L}$     & Unmarked layout produced by $\mathcal{F}$ \\
$\mathcal{L}^*, \mathcal{\hat{L}}$   & Watermarked layout, suspect layout \\
$o$, $id(o)$      & Candidate watermark object and stable identifier of $o$ \\
$n$         & Signal net \\
${E}_s$, $\mathrm{WM}_s$   & Eligible object set and watermark set for stage $s$ \\
$b_{tgt}$, $\pi_{tgt}$, $\hat{b}_P$, $\hat{b}_C$        & Target bit, target permutation index, and observed bit in placement and CTS \\
$r_s$             & Stage verification statistic\\
$r_P,r_C$         & Placement and CTS extraction rates, $r_P,r_C \in [0,1]$  \\
$r_R$           & Binary routing-stage result, $r_R \in \{0, 1\}$ \\
$\tau_P,\tau_C$ & Placement and CTS extraction-rate thresholds\\
$\alpha_R$ & Routing-stage false-positive probability budget\\
$\theta_{HPWL}$, $\delta_{guard}$ & Maximum allowable HPWL change and timing-slack degradation for placement edits \\
$R_{max}$ & Maximum Manhattan distance between eligible LCBs\\
$P_c$, $P_{c,s}$  & Combined and stage-level coincidence probabilities \\
$X_s$, $x_s$     &Number of evaluated and mismatched certified claims at $s\in\{P,C\}$ \\ 
$\Gamma_P,\Gamma_C$ & Serialized placement and CTS claim sets containing the stable identifier and target value of each accepted watermark object\\
$\Gamma$ & Watermark certificate, $\Gamma=(\Gamma_P,\Gamma_C)$ \\
$C_\Gamma$ & Encrypted and authenticated watermark certificate \\
$K_\Gamma$ & Certificate key derived from the secret key $K$ \\
$\mathcal{M}$     & PPA and legality metrics used for quality checks \\
$\operatorname{ID}(\mathcal D_0)$ &
SHA-256 identifier of the design version and watermark
configuration \\
$\nu$ &
Fresh public 96-bit nonce generated for each watermark instance and also used as the AES-GCM nonce for certificate encryption \\
$D_{\text{p}}$   & Maximum span for placement tuples \\
$q_R(n)$ & Per-net wrong-way fraction, $q_R(n)=w(n)/m(n)$ \\
$\varepsilon_m$   & Maximum allowable degradation budget for metric $m$ \\
$f$, $\lambda_{wm}$              & Routing watermark selection fraction; Wrong-way routing penalty strength \\
$\bar q_1$, $\bar q_0$ & Mean per-net wrong-way fractions of selected and unselected nets \\
$s_1^2$, $s_0^2$ & Sample variances of the per-net wrong-way fractions \\
$N_1$, $N_0$ & Numbers of selected and unselected eligible nets \\
$Z_R$, $p_R$      & Routing test statistic and corresponding p-value \\
$q_s$             & Fraction of eligible objects perturbed by an attacker\\
LCB               & Leaf clock buffer \\
PRF               & Pseudorandom function; HMAC-SHA256 in this work \\
AEAD & Authenticated encryption with associated data \\
carrier & PD property that encodes ownership evidence\\
hook  & Watermark insertion/verification point\\
claim  & One watermark observation used during verification\\
\hline
\end{tabular}
\end{table}

\subsection{Physical Design Flow}

A physical design (PD) flow $\mathcal{F}$ transforms a gate-level netlist into a legal
layout database. This work uses three key PD stages as watermark
carriers: placement, clock tree synthesis (CTS), and routing. These
stages are not independent. Placement determines the geometric
context for clock and signal routing; CTS introduces clock buffers
and fixes clock arrival relationships; routing realizes the final
interconnect and exposes detailed wire and via choices.

In this work, a \emph{watermark carrier} is a physical design
property or degree of freedom that can encode ownership evidence.
A \emph{watermark hook} is an insertion point in the physical design
flow at which \emph{PDMarks} performs watermark embedding or verification.
A physical design object \emph{carries} a watermark when its observed
state contributes to the ownership evidence checked by the verifier.




\subsection{Watermarking Requirements}

A PD watermark must satisfy several
requirements.

\textbf{Low PPA overhead.}
The marked layout must remain close to the owner's original
solution. The watermark should not introduce large timing,
power, area, or wirelength penalties.

\textbf{Sufficient capacity.}
The design must contain enough eligible objects to encode the
detectable signature. Capacity depends on the stage. 
Placement, CTS, and routing expose
different object populations; using all three stages increases the
number of independent observations available at verification. 

\textbf{Robust survival.}
The watermark must survive later PD steps. Placement
marks should not be destroyed by CTS or routing. CTS marks
should not be undone by routing optimization.
Routing marks must remain visible in the final routed
database. 

\textbf{Kerckhoffs security.}
The watermark objects must be hidden by the key, not by the
algorithm. If the source code is public, an adversary can
repeat every deterministic step that does not use the key.
Thus, watermark-object selection and target values must be derived
from a cryptographic PRF keyed by the owner. This is the
central design rule of our framework.

\subsection{Problem Formulation}
\label{subsec:problem_formulation}

The protected asset is the physical implementation produced by the
owner's PD flow. Let $\mathcal{D}_0$ denote the input design state,
including the netlist, technology, libraries, and implementation
constraints, and let
\begin{equation}
  \mathcal{L}=\mathcal{F}(\mathcal{D}_0)
\end{equation}
be the reference layout generated by the owner's flow $\mathcal{F}$.
Given a secret key $K$, the watermarking objective is to construct a
layout $\mathcal{L}^*$ that preserves circuit functionality and layout
legality, contains verifiable ownership evidence, and satisfies
\begin{equation}
  d_m(\mathcal{L}^*,\mathcal{L})\leq\varepsilon_m,
  \qquad \forall m\in\mathcal{M},
  \label{eq:ppa-bound}
\end{equation}
where $d_m$ measures the degradation of quality metric $m$, and
$\varepsilon_m$ is an owner-specified, design-dependent budget.
Acceptable budgets depend on the design, technology, and
implementation objectives, so we do not impose universal values.

At each available stage $s\in\mathcal{A}\subseteq\{P,C,R\}$, the
public procedure constructs an eligible object set $E_s$, and a
key-dependent procedure embeds stage-specific ownership evidence.
Verification produces a stage-level statistic $r_s$ with acceptance
threshold $\tau_s$. Ownership is accepted when at least two available
stages pass their corresponding tests:
\begin{equation}
  \left|
  \left\{
  s\in\mathcal{A}:r_s\geq\tau_s
  \right\}
  \right|\geq 2.
  \label{eq:ownership_rule}
\end{equation}
To prevent retrospective key selection, a claim is admissible only
when the claimed key opens an authenticated and time-stamped commitment
registered before release of the protected layout. The stage-specific
embedding procedures, watermark certificate, key commitment, and
verification protocol are described in Section~\ref{sec:methodology}.

\subsection{Attack Model}

We consider a white-box adversary who attempts to remove  
watermarks while preserving the value of the stolen implementation. The
adversary knows the source code, algorithms, Tcl scripts, command-line
options, tool logs, intermediate checkpoints, and integration points of
the watermarking hooks. The adversary may interrupt and resume the
flow, modify intermediate design databases, and rerun individual
PD stages with altered settings. The adversary
also has access to the same class of physical design tools as the IP
owner.

The adversary does not know the owner's secret key or the stage keys
derived from it. Consequently, even with full knowledge of the public
algorithm, the adversary does not know which objects are selected as watermark objects. We mainly consider two
attack categories: \emph{blind attack}, in which objects of a given type
are modified without knowing whether they are marked; and \emph{targeted
attack}, in which the adversary tries to infer watermark objects from
observable layout features before modifying them.
We also consider a false-claim adversary who searches for a key
that accidentally satisfies the public verification rule.
PDMarks prevents such post hoc key search by requiring the claimed
key to match a timestamped commitment created before the protected
design is released.

The framework protects the specific physical implementation generated
by the owner's flow. It does not prevent an attacker from independently
creating a different layout from the same RTL. 
Such a layout, even if comparable in PPA, 
does not preserve the stolen physical implementation. 



\section{Related Work}
\label{sec:related_work}

Watermarking has been studied at several levels of the
electronic design stack, including logic synthesis~\cite{Oliveira01}, FPGA
mapping~\cite{LachMSP98}, high-level synthesis~\cite{HongP98,QuP98},
and layout implementation~\cite{KahngMMP98,KahngLMM01}.
At higher abstraction levels, the watermark is often encoded
through constraints on resource binding, scheduling, logic
structure, or configuration bits. Such approaches protect the
functional or structural representation that is available at
that design level. They do not directly protect the tuned
physical implementation produced by a PD
flow. In contrast, our work treats the physical design
database itself as the protected asset.

Some watermarking methods embed signatures by adding
constraints to physical design optimization. Kahng et
al.~\cite{KahngMMP98,KahngLMM01} introduced watermarking
constraints for placement and routing. 
Their approach constrains selected standard cells to prescribed 
row parities and selected signal nets to bounded
wrong-way wiring. 
Sun et al.~\cite{SunGX06} encoded watermarks through
additional buffers inserted within the 
timing-feasible range of selected nets.
These methods established physical design watermarking as a
practical ownership mechanism. However, their security model
does not assume a public embedding algorithm. Under
Kerckhoffs, the added constraints or structural patterns can
become attack targets.
Cai et al.~\cite{CaiGBX07} embed ownership evidence in 
post-layout hard IP by scattering pseudorandom watermark 
regions and adding functionless polysilicon where needed 
to satisfy predefined layer-overlap constraints. 
However, an adversary can remove the watermark by rerunning 
polysilicon insertion without degrading design quality.
Recent work, including ICMarks~\cite{ZhangRPK25} and 
AutoMarks~\cite{ZhangRPK24,ZhangRPK25TODAES}, searches for a 
low-cost placement region using either a design-aware 
scoring function~\cite{ZhangRPK25} or a pretrained 
GNN~\cite{ZhangRPK24,ZhangRPK25TODAES}, and then constrains 
selected cells to remain within that region. 
These methods improve watermark fidelity and robustness 
against several practical attacks. However, the watermark 
remains localized in placement and may be exposed by 
watermark-related annotations stored in the design database. 
In addition, these methods do not protect later PD stages 
such as CTS and routing. 
\section{Methodology}
\label{sec:methodology}

\begin{figure*}
    \centering
\includegraphics[width=0.9\textwidth]{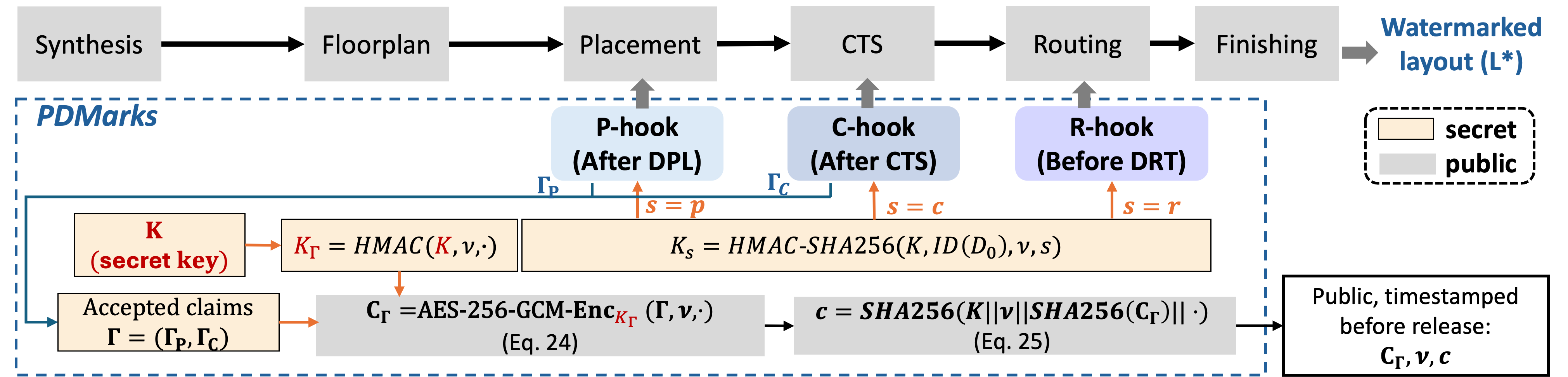}
  \caption{Overview of the \emph{PDMarks} framework. Owner-side embedding and ownership-proof construction.}
    \label{fig:framework}
    \vspace{-1.5em}
\end{figure*}
This section presents \emph{PDMarks}, a Kerckhoffs-compliant
watermarking framework that embeds ownership evidence during
placement, clock tree synthesis (CTS), and routing. Each stage uses a
different physical carrier and a stage-specific embedding procedure.
The public algorithm constructs the eligible objects, while a
secret-keyed pseudorandom function (PRF) determines the processing
order, selected objects, or target evidence according to the
corresponding stage.

Let $K$ denote the 32-byte secret key. 
A {\em watermark instance} is one
execution of {\em PDMarks} for a particular design version, producing one
watermarked layout, one encrypted certificate, and one timestamped
commitment. For each watermark instance, the owner generates a fresh
public 96-bit nonce $\nu$, 
which is used consistently across all
available watermarking stages, certificate encryption, and key
commitment for that instance.\footnote{The nonce $\nu$ distinguishes 
separate watermarking runs of the same design. A fresh $\nu$ is 
generated for every new watermark instance.}
For each stage
\(s\in\{P,C,R\}\), where \(P\), \(C\), and \(R\) denote placement,
CTS, and routing, respectively, \emph{PDMarks} derives a stage key
\begin{equation}
K_s =
\mathrm{HMAC\text{-}SHA256}(K, ID(\mathcal{D}_0),\nu,\texttt{stage}=s).
\label{eq:stage_key}
\end{equation}
All subsequent HMAC evaluations use explicit domain tags, such as
\texttt{tuple\_sort}, \texttt{bit}, \texttt{pair}, and \texttt{net},
to separate object selection, ordering, and target generation.
The design identifier $ID(\mathcal{D}_0)$ and
nonce $\nu$ bind the derived keys to a specific design version and watermark
instance.
As shown in Figure~\ref{fig:framework}, placement watermarking is
applied after detailed placement; CTS watermarking is applied after
clock-tree construction; and routing watermarking is applied before
detailed routing. Each (P-, C-, R-) hook performs bounded edits and
retains them only when the corresponding legality and quality checks
pass. The final accepted placement and CTS claims are recorded for
subsequent verification, while the routing watermark population is
reconstructed directly from the claimed key. No global
watermark-aware optimizer is required.

\subsection{Placement Watermarking}
\label{subsec:placement_wm}

Placement watermarking encodes ownership evidence in the relative
left-to-right order of nearby standard cells in the same row.
The mechanism is applied after detailed placement, when row legality
and local cell neighborhoods are available.

\textbf{Carrier.}
For a two-cell tuple \(o=(A,B)\) in the same row, the placement
watermark bit is
\begin{equation}
b_P(o) = \mathbbm{1}\{x(A) > x(B)\},
\label{eq:placement_pair_bit}
\end{equation}
where \(x(A)\) and \(x(B)\) are the bottom-left \(x\)-coordinates of
cells \(A\) and \(B\), respectively. Thus, \(b_P(o)=0\) when \(A\) is
to the left of \(B\), and \(b_P(o)=1\) otherwise.
For a three-cell tuple \(o=(A,B,C)\), the encoded value is the index of
the observed left-to-right permutation of \((A,B,C)\). A fixed public
lexicographic mapping assigns the six possible permutations to indices
\(\{0,1,\ldots,5\}\). Relative ordering is used instead of absolute
coordinates, so small legalization shifts do not change the extracted
value.

\textbf{Eligibility.}
Algorithm~\ref{alg:placement_wm} summarizes the placement watermarking
flow. The tool first enumerates same-row cell pairs and triples whose
horizontal spans do not exceed \(D_{\mathrm{p}}\) (Line 2). 
Candidates that violate
the public guards on half-perimeter wirelength (HPWL) change
\(\theta_{\mathrm{HPWL}}\), timing slack \(\delta_{\mathrm{guard}}\),
or row legality are removed (Line 3). The remaining candidates are
ordered by the keyed HMAC value
$F_{K_P}(\texttt{tuple\_sort},id(o))$,
where \(id(o)\) is a stable identifier constructed from the
hierarchical instance names of the cells in tuple $o$.
The algorithm
then scans the candidates in this keyed order and greedily selects each
tuple whose cells do not overlap with any previously selected tuple.
The resulting maximal non-overlapping set forms the placement
watermark set \(WM_P\) (Lines 4-5).

\begin{figure}
    \centering
\includegraphics[width=0.42\textwidth]{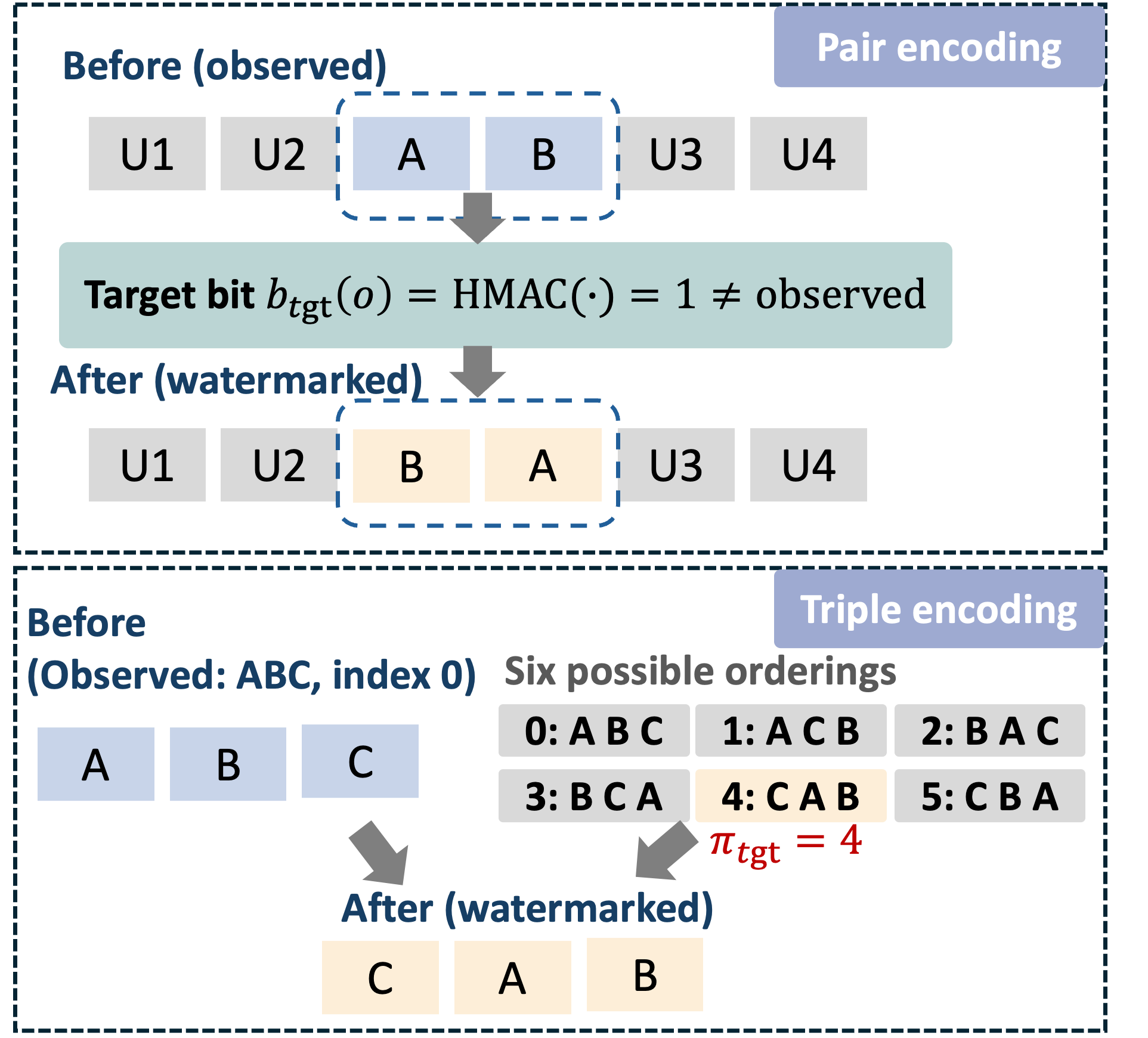}
  \caption{Placement watermarking by keyed local cell reordering.}
    \label{fig:place_wm}
    \vspace{-0.5em}
\end{figure}

\textbf{Embedding.}
For a selected pair \(o=(A,B)\), the target bit
$b_{\mathrm{tgt}}(o)$ is
\begin{equation}
b_{\mathrm{tgt}}(o)
=
\mathrm{HMAC}(K_P,\texttt{bit},id(o))[0] \,\&\, 1.
\label{eq:placement_target_bit}
\end{equation}
Here, \([0]\) denotes the first byte of the HMAC output and \(\&\,1\)
extracts its least significant bit. As shown in Figure~\ref{fig:place_wm}, 
if the current order of \(A\) and
\(B\) does not match \(b_{\mathrm{tgt}}(o)\), the two cells are
exchanged (Lines 7-9).
For a selected triple \(o=(A,B,C)\), the target permutation index is
\begin{equation}
\pi_{\mathrm{tgt}}(o)
=
\mathrm{HMAC}(K_P,\texttt{perm},id(o))[0:4] \bmod 6,
\label{eq:placement_target_perm}
\end{equation}
where \([0:4]\) denotes the first four bytes of the HMAC output
interpreted as an unsigned integer. As shown in Figure~\ref{fig:place_wm},
the three cells are reassigned to
their currently occupied sites according to $\pi_{\mathrm{tgt}}(o)$.
Since the edit reuses the tuple's existing sites, the
geometric displacement is local (Lines 11-12).
After legalization, {\em PDMarks} retains only 
the tuples whose target
ordering remains satisfied and 
records their stable identifiers and
target values in $\Gamma_P$ (Lines 15-18).

\begin{algorithm}[t]
\renewcommand{\algorithmicrequire}{\textbf{Input:}}
\renewcommand{\algorithmicensure}{\textbf{Output:}}
\caption{Placement Watermarking}
\label{alg:placement_wm}
\small
\begin{algorithmic}[1]
\REQUIRE Post-detailed-placement design \(O\); placement stage key \(K_P\);
 parameters \(D_{\mathrm{p}}\),
 $\theta_{HPWL}$, and \(\delta_{\mathrm{guard}}\)
\ENSURE Watermarked design \(O'\), placement watermark set \(WM_P\),
and placement certificate \(\Gamma_P\)
\STATE \(WM_P \leftarrow \emptyset\)
    \STATE \(U \leftarrow\) bounded same-row pairs and triples
    \STATE Filter \(U\) by HPWL, slack, and legality guards
    \STATE Order \(U\) by \(F_{K_P}(\texttt{tuple\_sort},id(\cdot))\)
    \STATE \(WM_P \leftarrow \mathrm{GreedyNonOverlap}(U)\)
\FOR{each tuple \(o \in WM_P\)}
    \IF{\(|o|=2\)}
        \STATE Compute \(b_{\mathrm{tgt}}(o)\) by Eq.~\eqref{eq:placement_target_bit}
        \STATE Swap the two cells if the observed bit differs from \(b_{\mathrm{tgt}}(o)\)
    \ELSE
        \STATE Compute \(\pi_{\mathrm{tgt}}(o)\) by Eq.~\eqref{eq:placement_target_perm}
        \STATE Reassign the three cells to their current sites according to
        \(\pi_{\mathrm{tgt}}(o)\)
    \ENDIF
\ENDFOR
\STATE Perform placement legalization
\STATE Remove from $WM_P$ any tuple whose target value is not preserved
\STATE Record the stable identifier and target value of 
each remaining tuple in \(\Gamma_P\)
\STATE \(O' \leftarrow O\)
\end{algorithmic}
\end{algorithm}


\textbf{Verification.}
The verifier authenticates and decrypts $C_\Gamma$ and obtains the 
placement claim set $\Gamma_P$. Each entry identifies a retained tuple and
its target value. The verifier locates the corresponding cells
in the suspect layout and extracts their observed ordering. 
Placement verification does not need to reconstruct the
embedding-time eligible set or repeat legalization and timing checks.
We define
\begin{equation}
\hat{y}_P(o,\hat{L})=
\begin{cases}
\hat{b}_P(o,\hat{L}), & |o|=2,\\
\hat{\pi}_P(o,\hat{L}), & |o|=3,
\end{cases}
\label{eq:placement_observed_value}
\end{equation}
where \(\hat{b}_P(o,\hat{L})\) is the observed pair-ordering bit and
\(\hat{\pi}_P(o,\hat{L})\) is the observed permutation index under the
fixed public permutation mapping.
The placement extraction rate is
\begin{equation}
r_P =
\frac{1}{|\Gamma_P|}
\sum_{(o,y_P(o))\in \Gamma_P}
\mathbbm{1}
\left\{
\hat{y}_P(o,\hat{L}) = y_P(o)
\right\}.
\label{eq:placement_extraction}
\end{equation}
A certified tuple that cannot be located contributes zero to the
indicator in Eq.~\eqref{eq:placement_extraction}.

\subsection{CTS Watermarking}
\label{subsec:cts_wm}

\begin{figure}
    \centering
\includegraphics[width=0.5\textwidth]{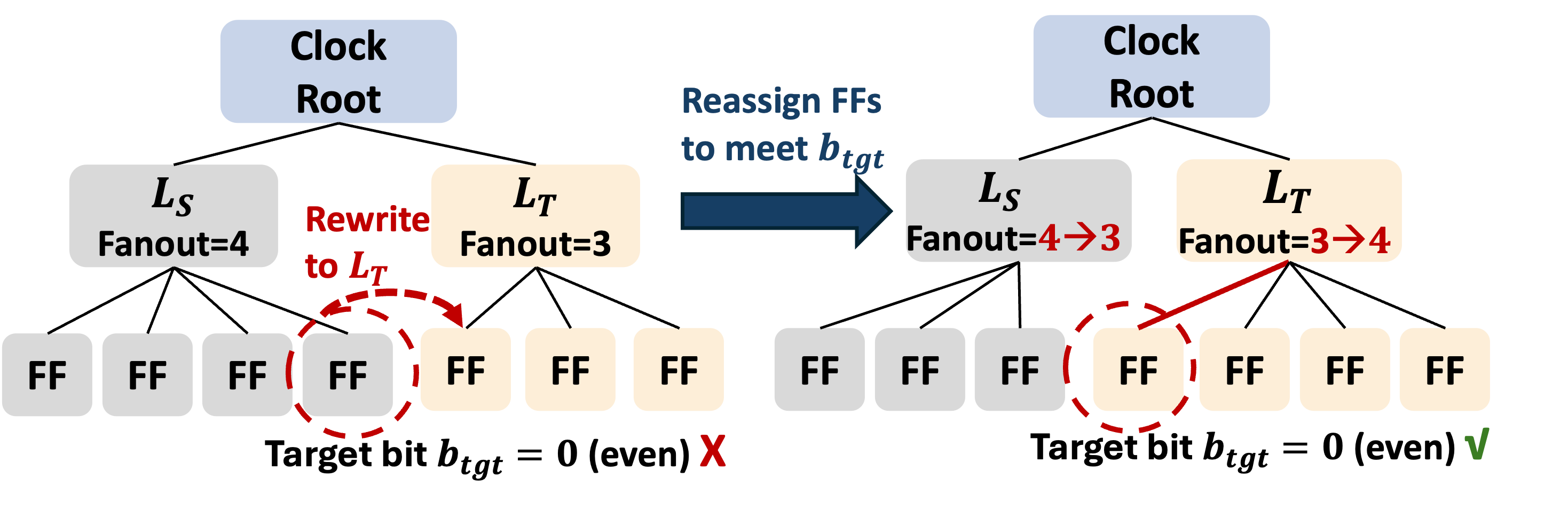}
  \caption{CTS watermarking by keyed LCB fanout-parity control.}
    \label{fig:cts_wm}
    \vspace{-1.5em}
\end{figure}

CTS watermarking embeds ownership evidence in the local fanout
structure of the clock tree after CTS has been completed. 
As shown in Figure~\ref{fig:cts_wm}, the carrier
is the parity of the sequential fanout of a key-selected leaf clock
buffer (LCB).

\textbf{Carrier.}
For an LCB \(L\), let \(\mathrm{seq\_fanout}(L)\) denote the number of
flip-flops (FFs) driven by \(L\). The CTS watermark bit is
\begin{equation}
b_C(L) = \mathrm{seq\_fanout}(L) \bmod 2.
\label{eq:cts_bit}
\end{equation}
Thus, \(b_C(L)=0\) when \(L\) drives an even number of sequential
sinks (FFs), and \(b_C(L)=1\) otherwise.
A CTS candidate is a compatible LCB pair \(o=(L_A,L_B)\). One LCB in
the pair is selected as the target LCB \(L_T\), which carries the
watermark claim. The other LCB, denoted \(L_S\), serves as the source
LCB for a possible boundary flip-flop reassignment.

\textbf{Eligibility.}
The CTS watermarking algorithm is shown in Alg.~\ref{alg:cts_wm}.
The eligible LCB pairs belong to the same
clock domain and have a Manhattan distance no
greater than $R_{max}$. This restriction
preserves clock intent and limits the electrical impact of any
reassignment (Line 2).
For each clock domain, 
the candidate pairs are ordered
by \(F_{K_C}(\texttt{pair\_sort},id(o))\) (Line 3),
where the identifier of an LCB is 
formed by concatenating its clock
domain and hierarchical buffer-instance name using a fixed,
unambiguous format.
The identifier of a pair $id(o)$ is formed from 
the lexicographically sorted LCB identifiers.
These identifiers are fixed before watermark embedding and
therefore do not change when a sequential sink is reassigned.

\textbf{Embedding.}
The embedding process continues until it accumulates
$N_C^{\max}$ accepted claims or exhausts the eligible 
population (Lines 5-7).
For a candidate pair \(o=(L_A,L_B)\), \emph{PDMarks} computes
\begin{equation}
d_C(o)
=
\mathrm{HMAC}(K_C,\texttt{pair},id(o))[0].
\label{eq:cts_decision}
\end{equation}
The least significant bit of \(d_C(o)\) gives the target parity, and the
next bit determines the target LCB:
\begin{equation}
b_{\mathrm{tgt}}(o) = d_C(o) \,\&\, 1,
\label{eq:cts_target_bit}
\end{equation}
\begin{equation}
L_T =
\begin{cases}
L_A, & ((d_C(o) \gg 1) \,\&\, 1)=0,\\
L_B, & \text{otherwise}.
\end{cases}
\label{eq:cts_target_lcb}
\end{equation}
The other LCB in the pair is denoted \(L_S\).
If the current parity of \(L_T\) already equals
\(b_{\mathrm{tgt}}(o)\), the claim is accepted without modifying the
clock tree (Lines 9-13). 
Otherwise, the tool searches for a legal local
reassignment. It enumerates boundary flip-flops currently driven by
\(L_S\) and located close to \(L_T\). Each candidate flip-flop is
temporarily reassigned to the clock net of \(L_T\). The edit is
accepted only if it changes the parity of \(L_T\) to
\(b_{\mathrm{tgt}}(o)\) and satisfies slew, load-capacitance, 
and skew checks. If any check fails, the reassignment is
reverted..
An accepted CTS claim is stored as \((L_T,b_{\mathrm{tgt}})\)
and all remaining candidate
pairs containing $L_T$ are removed
(Lines 14-26). 
After embedding, the stable identifier and target parity of every
accepted target LCB are recorded in $\Gamma_C$ (Line 27).

\begin{algorithm}[t]
\renewcommand{\algorithmicrequire}{\textbf{Input:}}
\renewcommand{\algorithmicensure}{\textbf{Output:}}
\caption{CTS Watermarking}
\label{alg:cts_wm}
\small
\begin{algorithmic}[1]
\REQUIRE Post-CTS design \(O\); CTS stage key \(K_C\);
parameters \(R_{\max}\) and \(N_C^{\max}\)
\ENSURE Watermarked design \(O'\), CTS watermark set \(WM_C\),
and CTS certificate \(\Gamma_C\)
\STATE \(WM_C \leftarrow \emptyset\)
\STATE Construct eligible LCB pairs \(\Pi\) based on clock domain
and \(R_{\max}\)
\STATE Order each clock-domain pair list by
$F_{K_C}(\cdot)$

\FOR{each pair \(o=(L_A,L_B)\in\Pi\)}
    \IF{\(|WM_C| = N_C^{\max}\)}
        \STATE \textbf{break}
    \ENDIF

    \STATE Compute \(d_C(o)\), \(b_{\mathrm{tgt}}(o)\), \(L_T\),
    and \(L_S\) by
    Eqs.~\eqref{eq:cts_decision}--\eqref{eq:cts_target_lcb}

    \IF{\(\mathrm{seq\_fanout}(L_T)\bmod 2
    = b_{\mathrm{tgt}}(o)\)}
        \STATE Accept \((L_T,b_{\mathrm{tgt}}(o))\) and add it to
        \(WM_C\)
        \STATE Remove from \(\Pi\) all remaining pairs containing \(L_T\)
        \STATE \textbf{continue}
    \ENDIF

    \STATE \(F \leftarrow
    \mathrm{BoundaryFFs}(L_S \rightarrow L_T)\)
    \STATE Order \(F\) by distance to \(L_T\)

    \FOR{each \(ff\in F\)}
        \STATE Temporarily connect \(ff\) to \(L_T\)'s clock net

        \IF{the target parity is satisfied and the timing/electrical
        checks pass}
            \STATE Accept \((L_T,b_{\mathrm{tgt}}(o))\) and add it to
            \(WM_C\)
            \STATE Remove from \(\Pi\) all remaining pairs containing
            \(L_T\)
            \STATE \textbf{break}
        \ELSE
            \STATE Revert the reassignment
        \ENDIF
    \ENDFOR
\ENDFOR

\STATE Record the stable identifier and target parity of each claim in
\(WM_C\) in \(\Gamma_C\)
\STATE \(O' \leftarrow O\)
\end{algorithmic}
\end{algorithm}


\textbf{Verification.}
The verifier obtains the CTS claim set $\Gamma_C$ from the
authenticated certificate. For each certified claim, it locates the target
LCB in the suspect clock tree and extracts its fanout parity.
CTS verification does not need to reconstruct the
embedding-time pair ordering, reassignment search, or electrical feasibility
checks. The CTS extraction rate is
\begin{equation}
r_C =
\frac{1}{|\Gamma_C|}
\sum_{(L_T,b_{\mathrm{tgt}})\in \Gamma_C}
\mathbbm{1}
\left\{
\hat{b}_C(L_T,\hat{L}) = b_{\mathrm{tgt}}
\right\}.
\label{eq:cts_extraction}
\end{equation}
Here, $\hat{b}_C(L_T,\hat{L})$ is the observed sequential-fanout
parity of $L_T$. A certified target LCB that cannot be located contributes
zero to the indicator.

\subsection{Routing Watermarking}
\label{subsec:routing_wm}

Routing watermarking differs from placement and CTS watermarking. It
does not assign an exact target bit to each selected net. Instead,
following the routing-constraint insight of~\cite{KahngMMP98},
\emph{PDMarks} selects a keyed population of signal nets and biases 
detailed routing to reduce wrong-way routing on that population.
The watermark is therefore detected statistically.

\begin{figure}
    \centering
\includegraphics[width=0.5\textwidth]{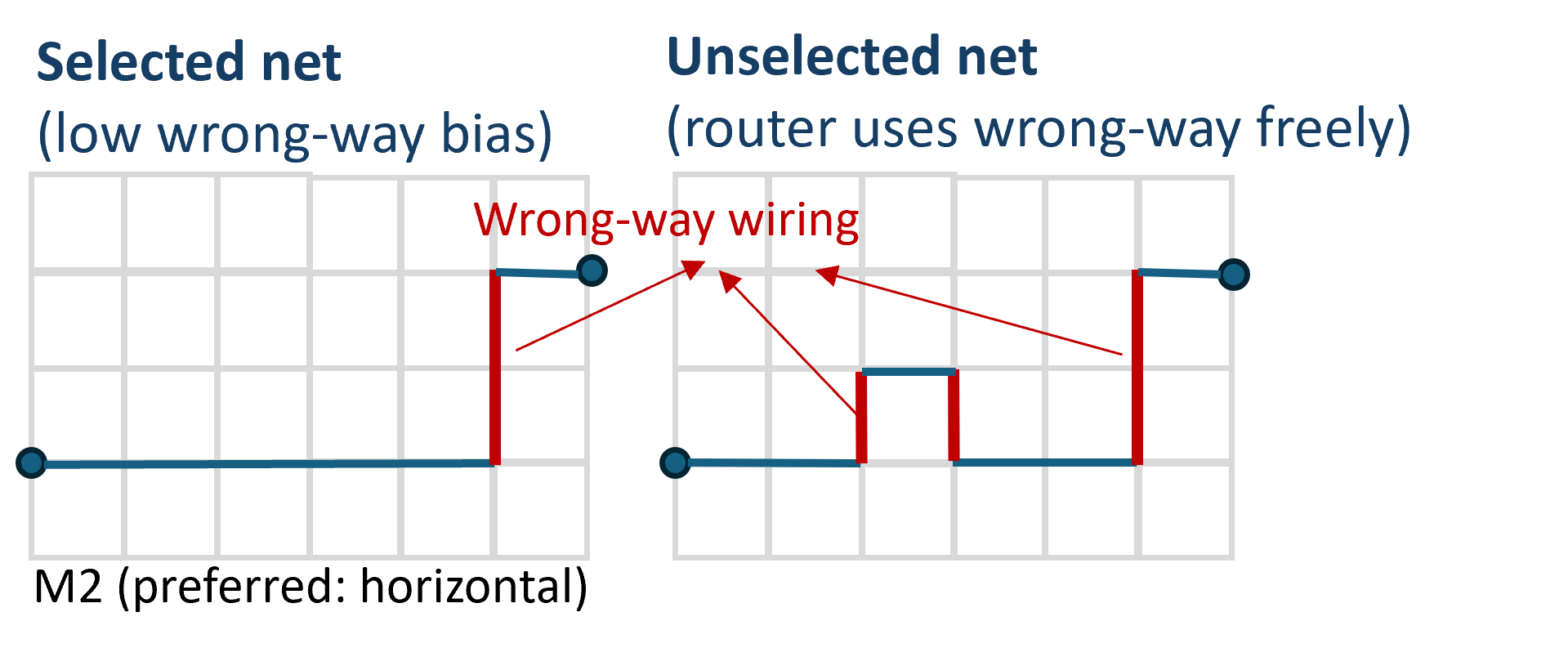}
  \caption{Routing watermarking by keyed wrong-way bias.}
    \label{fig:route_wm}
\end{figure}

\textbf{Carrier.}
Each routing layer has a preferred routing direction. 
As shown in Figure~\ref{fig:route_wm}, a routed segment
is called wrong-way if its direction is orthogonal to the preferred
direction of its layer. For a signal net \(n\), let \(w(n)\) be the
number of wrong-way routed segments and let \(m(n)\) be the total
number of routed segments. The per-net wrong-way fraction 
is computed as $q_R(n)=w(n)/m(n)$.

The routing watermark carrier is the aggregate wrong-way fraction of a
key-selected net population. This population-level carrier avoids
requiring any fixed route shape. The detailed router remains free to
choose legal routes, but selected nets are discouraged from using
wrong-way segments.

\textbf{Eligibility.}
The routing watermarking algorithm is shown in Alg.~\ref{alg:routing_wm}.
The eligible routing set \(E_R\) consists of signal nets in the
pre-detailed-routing database. Clock nets, along with power and 
ground nets, are excluded (Line 1). For each
eligible net \(n\), \emph{PDMarks} computes
\begin{equation}
u_R(n)
=
\frac{
\mathrm{HMAC}(K_R,\texttt{net},id(n))[0:4]
}{
2^{32}
}.
\label{eq:routing_score}
\end{equation}
The routing watermark set is
\begin{equation}
WM_R = \{n\in E_R : u_R(n) < f\},
\label{eq:routing_selection}
\end{equation}
where \(f\) is the routing watermark fraction. Each eligible net is
therefore selected with probability \(f\) (Lines 2-8).

\textbf{Embedding.}
During detailed routing, \emph{PDMarks} increases 
the cost of wrong-way
routing on nets in \(WM_R\). The strength of this penalty is controlled
by \(\lambda_{\mathrm{wm}}\) (Line 9). The router still enforces connectivity,
design rules, pin access, and congestion constraints. A selected net
may still contain wrong-way segments when they are required for legal
or high-quality routing. Thus, the routing watermarking biases the selected
population, but does not impose a fixed routing pattern.

\begin{algorithm}[t]
\renewcommand{\algorithmicrequire}{\textbf{Input:}}
\renewcommand{\algorithmicensure}{\textbf{Output:}}
\caption{Routing Watermarking}
\label{alg:routing_wm}
\small
\begin{algorithmic}[1]
\REQUIRE Pre-detailed-routing design \(O\); routing stage key \(K_R\);
watermark fraction \(f\); wrong-way penalty strength \(\lambda_{\mathrm{wm}}\)
\ENSURE Routed design and routing watermark set \(WM_R\)
\STATE \(E_R \leftarrow\) eligible signal nets in \(O\); \(WM_R \leftarrow \emptyset\)
\FOR{each net \(n \in E_R\)}
    \STATE Compute \(u_R(n)\) by Eq.~\eqref{eq:routing_score}
    \IF{\(u_R(n) < f\)}
        \STATE Regard \(n\) as watermarked net
        \STATE \(WM_R \leftarrow WM_R \cup \{n\}\)
    \ENDIF
\ENDFOR
\STATE Run detailed routing with wrong-way penalty \(\lambda_{\mathrm{wm}}\)
on nets in \(WM_R\)
\end{algorithmic}
\end{algorithm}

\textbf{Verification.}
After detailed routing, the verifier reconstructs \(WM_R\) from
the claimed key and computes \(q_R(n)\) for every eligible net. Let
\begin{equation}
\bar q_1=\frac{1}{N_1}\sum_{n\in WM_R}q_R(n),
\qquad
\bar q_0=\frac{1}{N_0}\sum_{n\in E_R\setminus WM_R}q_R(n),
\label{eq:routing_means}
\end{equation}
where \(N_1=|WM_R|\) and
\(N_0=|E_R\setminus WM_R|\). Since embedding penalizes wrong-way
routing only on selected nets, the expected watermark signature is
\(\bar q_1<\bar q_0\). 
Verification uses the one-sided hypothesis test
\begin{equation}
H_0:\mu_1\ge\mu_0,
\qquad
H_1:\mu_1<\mu_0,
\label{eq:routing_hypothesis}
\end{equation}
where $\mu_1$ and $\mu_0$ are the population means of the
per-net wrong-way fractions for selected and unselected nets.
 Let
$s_1^2$ and $s_0^2$ be the corresponding sample variances. We use
the large-sample Welch statistic
\begin{equation}
Z_R=
\frac{\bar q_0-\bar q_1}
{\sqrt{s_1^2/N_1+s_0^2/N_0}}.
\label{eq:routing_z}
\end{equation}
A large positive value of $Z_R$ indicates that selected nets use
wrong-way routing less frequently than unselected nets. The one-sided
routing $p$-value is
\begin{equation}
p_R = 1-\Phi(Z_R),
\label{eq:routing_pvalue}
\end{equation}
where \(\Phi(\cdot)\) is the standard normal cumulative distribution
function. The routing stage passes if
\begin{equation}
p_R \le \alpha_R,
\label{eq:routing_pass}
\end{equation}
where \(\alpha_R\) is the routing-stage false-positive budget. For the
combined ownership test, we define
\begin{equation}
r_R =
\mathbbm{1}\{p_R \le \alpha_R\}.
\label{eq:routing_extraction}
\end{equation}

\subsection{Certificate and Ownership Verification}
\label{subsec:ownership_verification}

\textbf{Certificate.}
After embedding, the final accepted placement and CTS claims are
recorded in $\Gamma=(\Gamma_P,\Gamma_C)$. The placement certificate
$\Gamma_P$ contains the stable identifier and target ordering of each
cell tuple, while $\Gamma_C$ contains the stable identifier
and target parity of each target LCB. 
The certificate is serialized using an unambiguous 
encoding and encrypted with AES-256-GCM.\footnote{For every multi-field
cryptographic input, each field is prefixed with its byte length
before concatenation to prevent ambiguous serialization.} 
We derive
\begin{equation}
\begin{aligned}
K_\Gamma &= \operatorname{HMAC\text{-}SHA256}
(K, ID(\mathcal{D}_0), \nu, \texttt{cert}),\\
C_\Gamma &= \operatorname{AES-256-GCM\text{-}Enc}_{K_\Gamma}(\Gamma, ID(\mathcal{D}_0), \nu).
\end{aligned}
\label{eq:certificate}
\end{equation}

Here, $C_\Gamma$ contains both the encrypted certificate and the
AES-GCM authentication tag. The nonce $\nu$ was generated once before
embedding (see Sec.~\ref{sec:methodology}) and is reused here; 
it is not regenerated during certificate construction.
$C_\Gamma$ can be stored publicly and 
introduces no
additional secret beyond $K$.

\textbf{Key commitment.}
Before releasing the protected layout, the owner reuses the public
watermark-instance nonce $\nu$ generated before embedding and
registers
\begin{equation}
c = \operatorname{SHA256}\!\left(
K \parallel \operatorname{ID}(\mathcal{D}_0) \parallel \nu
\parallel \operatorname{SHA256}(C_\Gamma)
\right).
\label{eq:key_commitment}
\end{equation}
Let
$R =
(\operatorname{ID}(\mathcal D_0), \nu,
 \operatorname{SHA256}(C_\Gamma), c).$
The owner submits the SHA-256 digest of $R$ in an RFC~3161
time-stamp request to an independent time-stamping authority
(e.g., FreeTSA~\cite{freetsa}).  
The authority returns a signed time-stamp
token proving that the signed record existed at the recorded time.
The earliest valid record is admissible, and later records
for the same owner and design identifier are ignored. This public record
binds one key and certificate to the design before layout release and
prevents retrospective key selection. 

\begin{figure}
    \centering
\includegraphics[width=0.5\textwidth]{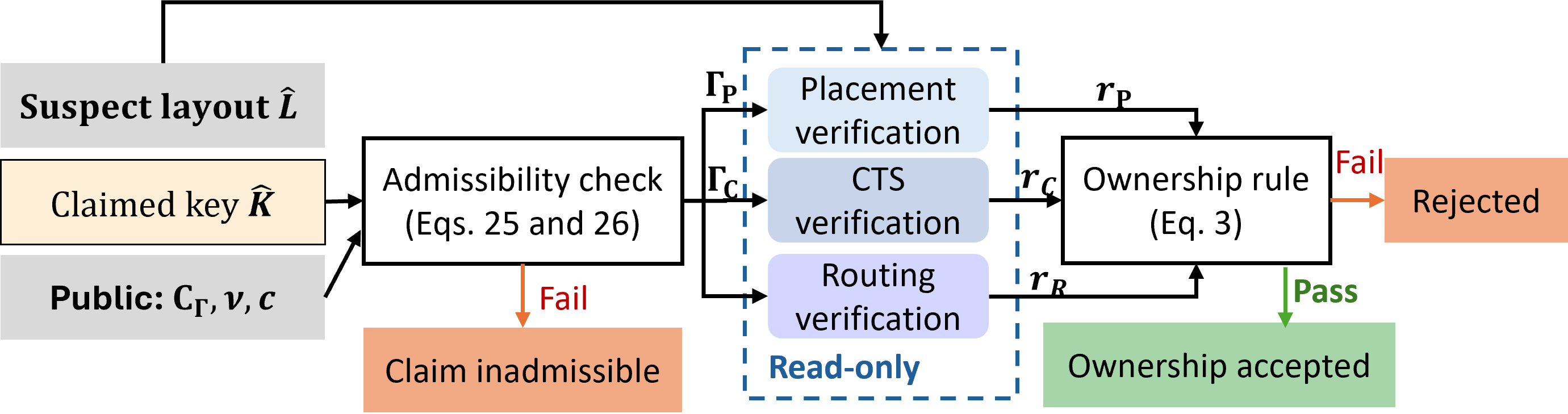}
  \caption{Read-only ownership verification of a suspect layout.}
    \label{fig:verification}
    \vspace{-1.5em}
\end{figure}

\textbf{Ownership verification.}
Figure~\ref{fig:verification} illustrates the verification
process. Given a suspect layout $\widehat{\mathcal{L}}$, a claimed key
$\widehat{K}$, and the public ciphertext $C_\Gamma$, the verifier first
checks whether the claimed key opens the registered commitment:
\begin{equation}
  \operatorname{SHA256}\!\left(
  \widehat{K}\parallel
  \operatorname{ID}(\mathcal{D}_0)\parallel\nu
  \parallel \operatorname{SHA256}(C_\Gamma)
  \right)=c.
\label{eq:commitment_check}
\end{equation}
A claim that fails this check is inadmissible. Otherwise, the verifier
derives
\begin{equation}
  \widehat{K}_\Gamma=
  \operatorname{HMAC\text{-}SHA256}
  (\widehat{K},ID(\mathcal{D}_0), \nu,\texttt{cert}).
\end{equation}
The verifier then authenticates and decrypts $C_\Gamma$ using the 
public nonce $\nu$ and $\operatorname{ID}(\mathcal D_0)$ 
as the associated data. Failed certificate
authentication also makes the claim inadmissible.

For placement and CTS, the verifier evaluates the fixed accepted
claims in $\Gamma_P$ and $\Gamma_C$, respectively. A certified object
that is missing or cannot be located unambiguously is counted as a
mismatch. For routing, the verifier derives $\widehat{K}_R$ using
Eq.~\eqref{eq:stage_key} and reconstructs the selected net population.
The resulting statistics $r_P$, $r_C$, and $r_R$ are computed using
Eqs.~\eqref{eq:placement_extraction}, \eqref{eq:cts_extraction}, and
\eqref{eq:routing_extraction}, respectively, and the ownership decision
follows Eq.~\eqref{eq:ownership_rule}. The thresholds are selected to
control the false-acceptance probability for a single committed key
and are evaluated in Section~\ref{subsec:wrong-key}.

\subsection{Security Discussion}
\label{subsec:security_discussion}

\emph{PDMarks} follows Kerckhoffs's principle because the eligible-set
construction, embedding mechanisms, and verification procedures
are public, while only the secret key $K$ remains secret.
Under the PRF assumption for HMAC-SHA256, the key-selected
watermark objects and their target values cannot be predicted
from the public algorithms. Domain-separated stage keys further
ensure that knowledge of the watermark selection at one stage
does not reveal the selections at other stages. Therefore, an
adversary without $K$ cannot directly identify the watermarked
objects and must instead perturb a broad eligible population
(i.e., {\em blind attack}) or
attempt to infer marked objects from observable physical
features (i.e., {\em targeted attack}). 
Section~\ref{sec:attack_evaluation} evaluates these attacks and quantifies 
their effects on watermark extraction and
implementation quality.
The public certificate and commitment procedures reveal no
accepted placement or CTS claims without $K$ and introduce no secret
beyond the secret key.

\subsection{Runtime Complexity and Scalability}
Let $N_{\mathrm{cell}}$ and $N_{\mathrm{LCB}}$ denote the numbers
of movable standard cells and leaf clock buffers, respectively,
and let $B_C$ denote the total number of boundary flip-flop
candidates examined during CTS embedding. \emph{PDMarks} introduces
lightweight watermark-specific computation around the underlying
physical design operations.

For \textbf{placement}, sorting cells by row and location requires
$O(N_{\mathrm{cell}}\log N_{\mathrm{cell}})$ time. With the tuple span
bounded by $D_{\mathrm{p}}$, candidate enumeration is local. Keyed
ordering and greedy non-overlap selection require
$O(|E_P|\log |E_P|)$ time. Placement verification builds a
stable-identifier index and checks the $|\Gamma_P|$ certified tuples,
requiring $O(N_{\mathrm{cell}}+|\Gamma_P|)$ time.
For \textbf{CTS}, direct enumeration of all LCB pairs has worst-case
complexity $O(N_{\mathrm{LCB}}^2)$, while clock-domain filtering and
the distance bound $R_{\max}$ substantially reduce the practical
candidate population. Keyed ordering requires
$O(|E_C|\log |E_C|)$ time, and pair processing requires
$O(|E_C|+B_C)$ time.
CTS verification requires
$O(N_{\mathrm{LCB}}+|\Gamma_C|)$ time.
For \textbf{routing}, watermark-net selection requires
$O(|E_R|)$ HMAC evaluations. The watermark adds only a constant-time
term to each routing-edge cost evaluation and therefore does not change
the asymptotic complexity of detailed routing. Routing verification
scans the eligible nets and routed segments once, requiring
$O\!\left(|E_R|+\sum_{n\in E_R}m(n)\right)$
time.

Overall, the bounded placement span and CTS distance threshold keep the
practical candidate populations manageable, while routing selection
and verification scale linearly with routed design size.

\section{Experiments}
\label{sec:experiments}

We evaluate \emph{PDMarks} in the setting where the design owner has
already obtained a physical implementation through a tuned PD flow,
which we call the \emph{reference flow}. This
unmarked implementation represents the owner's PD IP and is denoted as
the \emph{reference layout}. The owner then enables the \emph{PDMarks} hooks in
the same PD flow to produce a \emph{watermarked layout}. The reference and
watermarked runs use the same synthesized netlist, floorplan, timing
constraints, technology files, routing-layer configuration, and tool
options. The only intentional difference is the insertion of watermarking
hooks at placement, CTS, and routing.
Our evaluation addresses the following questions.
\begin{itemize}[noitemsep,topsep=0pt,leftmargin=*]
\item Does the physical design database provide sufficient 
watermark capacity at placement, CTS, and routing stages? 
\item What PPA overheads are introduced when watermarking is
enabled in the owner's tuned PD flow?
\item How does \emph{PDMarks} compare with prior PD
watermarking strategies under the same implementation flow? 
\item Can the selected watermark objects
be extracted and verified after the complete PD flow? 
\item Can ownership claims be forged without the secret key?

\end{itemize}

\subsection{Experimental Setup}
We integrate the embedding and verification hooks of \emph{PDMarks} into
OpenROAD-flow-scripts (ORFS); see the GitHub
repo~\cite{PDMarks_repo}.
The default ORFS flow is used as the reference flow. 
All comparisons are made
between the reference layout and the corresponding watermarked layout at
the same flow checkpoint.
We evaluate \emph{PDMarks} using four testcases 
\texttt{JPEG}, \texttt{SweRV}, \texttt{Ariane}, and
\texttt{BP} in NanGate45 and four testcases
\texttt{JPEG}, \texttt{SweRV}, \texttt{Ariane}, and
\texttt{CVA6} in ASAP7;
these are publicly available in the OpenROAD~\cite{OpenROAD} 
and MacroPlacement~\cite{MacroPlacement} 
GitHub repositories. 
Final PPA metrics are collected after
detailed routing. We report routed
wirelength (rWL), worst negative slack (WNS), total
negative slack (TNS), and total power.
Table~\ref{tab:exp_setup} summarizes the
characteristics of these testcases and the PPA results
of the reference flow.\footnote{{\em PDMarks} runtime overheads 
for placement, CTS and routing total between 
3\% and 9\% of the reference flow runtimes in Table II.}

\textbf{Parameters}. Unless otherwise stated, all watermarking 
runs use a 32-byte secret key.
The stage keys are derived as described in
Section~\ref{sec:methodology}. 
Since ASAP7 cells, sites, and tracks are substantially 
smaller than those of NanGate45, we use different 
parameter values with these two technology nodes.
For NanGate45, 
the default placement parameters are
$D_{\mathrm{p}}=2\,\mu\mathrm{m}$,
$\theta_{\mathrm{HPWL}}=100$ DBU, and
$\delta_{\mathrm{guard}}=0.02\,\mathrm{ns}$. 
On ASAP7, $D_{\mathrm{p}}=1\mu\mathrm{m}$,
$\theta_{\mathrm{HPWL}}=100$ DBU, and
$\delta_{\mathrm{guard}}=0.01\mathrm{ns}$. 
For CTS, two leaf clock
buffers form an eligible pair when their centroids lie within a
bounded Manhattan distance $R_{\max}$. The default is
$R_{\max}=50\,\mu\mathrm{m}$ for NanGate45 and 
$R_{\max}=14\,\mu\mathrm{m}$ for ASAP7.
We set the CTS claim budget to $N_C^{\max}=24$ for all experiments.
All evaluated designs provide enough feasible candidates to reach this
budget.
The default
routing watermark fraction and wrong-way penalty strength are
$f=0.02$ and $\lambda_{\mathrm{wm}}=100$.
Watermarking sensitivity to these parameters is studied in
Section~\ref{subsec:param_sensitivity}.

\textbf{Baselines.}
We compare \emph{PDMarks} with Row-parity~\cite{KahngMMP98,KahngLMM01},
Buffer-insertion~\cite{SunGX06}, and ICMarks~\cite{ZhangRPK25}.
Row-parity encodes ownership by constraining selected cells to rows
with prescribed parity. Buffer-insertion encodes signature bits
through the number of buffers inserted on selected nets within their
timing-feasible ranges. ICMarks searches for a low-cost placement
region and constrains selected cells to remain within that region.
We do not include Cell-scattering~\cite{CaiGBX07} because its
post-layout polysilicon watermark can be removed by rerunning
polysilicon insertion without degrading design quality.
AutoMarks~\cite{ZhangRPK24,ZhangRPK25TODAES} uses a pretrained GNN
to accelerate the region search of ICMarks but applies a similar
region-based placement watermark. We therefore use ICMarks as the
representative baseline for this class of methods.

All baseline methods are evaluated using the same synthesized netlist,
floorplan, timing constraints, technology files, routing configuration,
and downstream ORFS flow as \emph{PDMarks}. Each baseline is inserted
at the stage prescribed by its original method, and final PPA metrics
are collected after detailed routing. 
We use the watermark sizes and parameter settings
recommended in the original papers.


\begin{table*}[t]
\caption{Benchmark characteristics and reference flow results.}
\label{tab:exp_setup}
\centering
\scriptsize
\resizebox{0.9\textwidth}{!}{%
\begin{tabular}{|c|l|c|c|c|c|c|c|c|c|c|c|}
\hline
Platform               & Design & \#Std cells & \#Macros & \#Nets & TCP (ns) & Util (\%) & WNS (ns) & TNS (ns) & rWL (um) & Power (mw) & Runtime (s) \\ \hline
\multirow{4}{*}{NG45}  & JPEG & 60,632 & 0 & 71,673 & 1.00 & 85 & -0.141 & -35.433 & 581,986 & 592 & 1,433 \\ \cline{2-12}
                       & SweRV & 102,834 & 28 & 109,728 & 2.00 & 60 & -0.283 & -365.187 & 3,844,113 & 274 & 5,279 \\ \cline{2-12}
                       & Ariane & 190,249 & 136 & 201,413 & 3.50 & 36 & -0.028 & -0.524 & 7,688,658 & 254 & 7,978 \\ \cline{2-12}
                       & BP & 96,085 & 26 & 105,799 & 4.80 & 60 & -0.461 & -0.461 & 3,174,283 & 141 & 2,067 \\ \hline
\multirow{4}{*}{ASAP7}  & JPEG & 65,038 & 0 & 71,522 & 0.54 & 70 & -0.020 & -0.940 & 161,844 & 175 & 2,100 \\ \cline{2-12}
                       & SweRV & 136,208 & 28 & 115,596 & 1.46 & 30 & -0.065 & -3.338 & 1,133,210 & 94 & 12,687 \\ \cline{2-12}
                       & CVA6 & 126,625 & 8 & 128,113 & 0.95 & 70 & -0.043 & -0.983 & 667,566 & 162 & 5,525 \\ \cline{2-12}
                       & Ariane & 224,380 & 37 & 210,946 & 1.50 & 70 & 0.003 & 0.000 & 1,435,591 & 129 & 9,220 \\ \hline
\end{tabular}}
\vspace{-1em}
\end{table*}

\subsection{Watermarking Capacity}
\label{subsec:capacity}

We first evaluate whether the physical design database has
enough eligible objects to carry a watermark at each stage.
Table~\ref{tab:capacity} reports the population of publicly eligible carriers 
${E}_s$ and the selected watermark population
$WM_s$. The set ${E}_s$ is determined by the public
structural rules of the corresponding stage, while $WM_s$ is
the subset selected by the stage key. Thus, 
$|{E}_s|$ measures the available design
degrees of freedom, and $|WM_s|$ measures the number of
claims carried by the watermarked implementation.
For placement and CTS, the selected counts in
Table~\ref{tab:capacity} are the final accepted claims after feasibility
checks, so $|WM_P|=|\Gamma_P|$ and $|WM_C|=|\Gamma_C|$.

Table~\ref{tab:capacity} shows that all three stages provide sufficient capacity for
ownership verification. The selected populations are substantially
smaller than the corresponding populations of publicly eligible carriers, which is
important for Kerckhoffs security: the public construction identifies
the class of possible carriers, but the secret key determines which
objects actually contribute to the ownership proof.
Note that we do not apply routing watermarking on ASAP7,
because wrong-way routing is disallowed by the routing rules
used in this technology node.
This illustrates the intended role of multi-stage
watermarking: the framework does not depend on a single physical carrier,
and verification can be based on the stages that expose usable degrees of
freedom in a given technology and flow.

\begin{table}[t]
\caption{Watermarking capacity.}
\label{tab:capacity}
\centering
\resizebox{0.4\textwidth}{!}{%
\begin{tabular}{|c|l|c|c|}
\hline
Stage & Design & Eligible $|{E}_s|$ & Selected $|WM_s|$ \\ \hline
\multirow{8}{*}{Placement} & JPEG (NG45) & 704 & 128 \\ \cline{2-4}
                            & SweRV (NG45) & 528 & 89 \\ \cline{2-4}
                            & Ariane (NG45) & 883 & 124 \\ \cline{2-4}
                            & BP (NG45) & 251 & 49 \\ \cline{2-4}
                            & JPEG (ASAP7) & 495 & 83 \\ \cline{2-4}
                            & SweRV (ASAP7) & 606 & 93 \\ \cline{2-4}
                            & CVA6 (ASAP7) & 1152 & 170 \\ \cline{2-4}
                            & Ariane (ASAP7) & 822 & 131 \\ \hline
\hline
Stage & Design & Eligible $|{E}_s|$ & Selected $|\mathcal{WM}_s|$ \\ \hline
\multirow{8}{*}{CTS} & JPEG (NG45) & 8620 & 24 \\ \cline{2-4}
                            & SweRV (NG45) & 17950 & 24 \\ \cline{2-4}
                            & Ariane (NG45) & 36024 & 24 \\ \cline{2-4}
                            & BP (NG45) & 29224 & 24 \\ \cline{2-4}
                            & JPEG (ASAP7) & 1471 & 24 \\ \cline{2-4}
                            & SweRV (ASAP7) & 2815 & 24 \\ \cline{2-4}
                            & CVA6 (ASAP7) & 1463 & 24 \\ \cline{2-4}
                            & Ariane (ASAP7) & 4228 & 24 \\ \hline
\hline
Stage & Design & Eligible $|{E}_s|$ & Selected $|\mathcal{WM}_s|$ \\ \hline
\multirow{4}{*}{Routing} & JPEG (NG45) & 70877 & 1417 \\ \cline{2-4}
                            & SweRV (NG45) & 107577 & 2151 \\ \cline{2-4}
                            & Ariane (NG45) & 197271 & 3945 \\ \cline{2-4}
                            & BP (NG45) & 102707 & 2054 \\ \cline{2-4}\hline
\end{tabular}
}
\vspace{-1em}
\end{table}

\subsection{PPA Overheads and Comparison with Prior Works}
\label{subsec:ppa}

Next, we measure the implementation 
cost of watermarking and the
statistical strength of the resulting 
watermark evidence.
We also compare \emph{PDMarks} with 
representative prior PD watermarking methods.

For \emph{PDMarks}, we evaluate four flows, namely, 
placement-only, CTS-only, routing-only, and
all-stage watermarking. 
The single-stage flows are ablation studies used to quantify the
individual cost and evidence strength of each carrier. The full
ownership rule in Eq.~\ref{eq:ownership_rule} is applied 
only to the all-stage flow.
Table~\ref{tab:ppa} reports WNS, TNS,
total power, routed wirelength, and the coincidence
probability $P_c$ for NanGate45 and ASAP7.

For each PPA metric, we report the
signed change $\Delta m$ with respect to the reference layout, i.e.,
\begin{equation}
\Delta m = m(\mathcal{L}^{*}) - m(\mathcal{L})
\end{equation}
where $\mathcal{L}$ is the reference layout and $\mathcal{L}^{*}$ is
the corresponding watermarked layout. 

The coincidence probability $P_c$ characterizes the statistical
strength of the observed watermark evidence. Specifically, 
it measures the probability that an unwatermarked implementation
produces evidence as strong as the watermark evidence purely
by coincidence. A smaller $P_c$ therefore indicates that 
the observed watermark is less likely to arise accidentally.
Following~\cite{KahngMMP98},
we model each watermark claim as a Bernoulli trial that is satisfied
with probability $p_s$ under a random (unwatermarked) implementation.
Let $X_s = |\widehat{\mathrm{WM}}_s|$ be the number of verified
claims in stage $s$, and let $x_s$ be the number of mismatched
claims observed in the suspect layout. The stage-level coincidence
probability is
\begin{equation}
P_{c,s} =
\sum_{i=0}^{x_s}
\binom{X_s}{i}
(1 - p_s)^i \, p_s^{\,X_s - i}.
\label{eq:pc_stage}
\end{equation}
For placement and CTS, we use $p_s=0.5$, 
corresponding to the conservative case 
in which each claim is satisfied with 
probability $1/2$ in an unwatermarked 
layout.\footnote{For placement, this assumes 
that all claims are pair-based. 
Triple-based claims have a lower 
random-match probability of $1/6$, 
so using $p_P=0.5$ gives a 
conservative upper bound on the 
placement coincidence probability.}
For routing, 
$P_{c,R}$ is given by the net-level
one-sided $p$-value in Eq.~\ref{eq:routing_pvalue}.
To provide a compact summary of the multi-stage evidence,
we define the approximate joint all-stage coincidence probability
\begin{equation}
P_c = \prod_{s \in \mathcal{A}} P_{c,s}.
\label{eq:pc_total}
\end{equation}
This calculation assumes independence among the 
stage-level coincidence events. It summarizes the strength of 
the complete observed evidence.
The calculation of $P_c$ for baseline methods 
is detailed in our GitHub repository~\cite{PDMarks_repo}.

\begin{figure}
    \centering
    \includegraphics[width=0.49\textwidth]{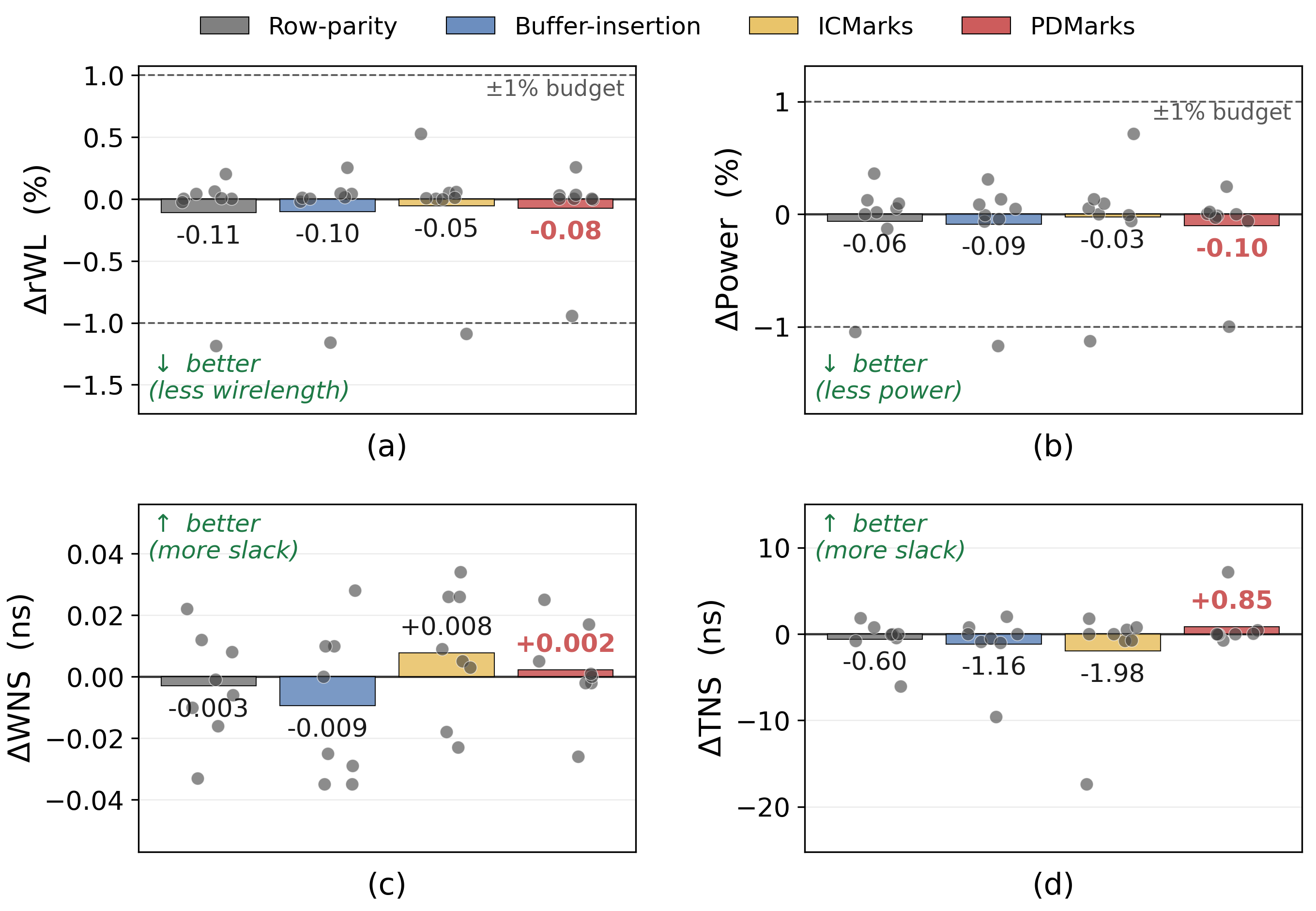}
  \caption{PPA changes of \emph{PDMarks} and baseline methods. }
    \label{fig:ppa}
\vspace{-1em}
\end{figure}

\begin{table}[t]
\caption{Comparison of quality and security among different methods in NanGate45
and ASAP7. rWL in $\mu$m, power in mW, and WNS and TNS in ns. Best results in blue font.}
\label{tab:ppa}
\centering
\footnotesize
\resizebox{0.5\textwidth}{!}{%
\begin{tabular}{|l|ll|l|l|l|l|l|}
\hline
Design                                                                 & \multicolumn{2}{l|}{Method}                               & $\Delta$rWL & $\Delta$Power & $\Delta$WNS & $\Delta$TNS & $P_c$ \\ \hline
\multirow{7}{*}{\begin{tabular}[c]{@{}l@{}}JPEG\\ NG45\end{tabular}} & \multicolumn{2}{l|}{Row-parity} & -6,901 & -6.193 & -0.033 & -0.811 & $\mathrm{2.8e-17}$ \\ \cline{2-8}
                                                                       & \multicolumn{2}{l|}{Buffer-insertion} & -6,731 & -6.922 & -0.035 & -0.899 & $\mathrm{1.5e-20}$ \\ \cline{2-8}
                                                                       & \multicolumn{2}{l|}{ICMarks} & -6,332 & -6.677 & 0.034 & -0.783 & $\mathrm{1.3e-22}$ \\ \cline{2-8}
                                                                       & \multicolumn{1}{l|}{\multirow{4}{*}{PDMarks}} & P-only & -5,645 & -5.954 & 0.027 & -0.142 & $\mathrm{2.9e-39}$ \\ \cline{3-8}
                                                                       & \multicolumn{1}{l|}{}                         & C-only & -3,890 & -4.033 & 0.016 & -0.487 & $\mathrm{6.0e-8}$ \\ \cline{3-8}
                                                                       & \multicolumn{1}{l|}{}                         & R-only & 18 & 0.055 & -0.002 & 0.023 & $\mathrm{3.5e-198}$ \\ \cline{3-8}
                                                                       & \multicolumn{1}{l|}{}                         & All-stage & -5,494 & -5.901 & 0.025 & -0.741 & \blue{\bf $\mathbf{4.8e-235}$} \\ \hline
\multirow{7}{*}{\begin{tabular}[c]{@{}l@{}}SweRV\\ NG45\end{tabular}} & \multicolumn{2}{l|}{Row-parity} & 1,616 & -0.350 & -0.016 & -6.067 & $\mathrm{9.1e-13}$ \\ \cline{2-8}
                                                                       & \multicolumn{2}{l|}{Buffer-insertion} & 1,623 & -0.178 & -0.029 & -9.615 & $\mathrm{4.3e-19}$ \\ \cline{2-8}
                                                                       & \multicolumn{2}{l|}{ICMarks} & 1,976 & -0.165 & -0.018 & -17.402 & $\mathrm{2.8e-17}$ \\ \cline{2-8}
                                                                       & \multicolumn{1}{l|}{\multirow{4}{*}{PDMarks}} & P-only & 703 & 0.017 & -0.012 & 8.223 & $\mathrm{1.6e-27}$ \\ \cline{3-8}
                                                                       & \multicolumn{1}{l|}{}                         & C-only & 2,658 & -0.226 & 0.003 & -2.124 & $\mathrm{6.0e-8}$ \\ \cline{3-8}
                                                                       & \multicolumn{1}{l|}{}                         & R-only & 171 & 0.009 & -0.001 & -0.402 & $\mathrm{<1e-300}$ \\ \cline{3-8}
                                                                       & \multicolumn{1}{l|}{}                         & All-stage & 1,146 & -0.161 & -0.026 & 7.156 & \blue{$\mathbf{<1e-300}$} \\ \hline
\multirow{7}{*}{\begin{tabular}[c]{@{}l@{}}Ariane\\ NG45\end{tabular}} & \multicolumn{2}{l|}{Row-parity} & 265 & 0.050 & -0.010 & -0.457 & $\mathrm{5.4e-20}$ \\ \cline{2-8}
                                                                       & \multicolumn{2}{l|}{Buffer-insertion} & 1,244 & -0.107 & -0.025 & -0.518 & $\mathrm{4.9e-32}$ \\ \cline{2-8}
                                                                       & \multicolumn{2}{l|}{ICMarks} & 151 & -0.020 & 0.026 & 0.517 & $\mathrm{6.5e-27}$ \\ \cline{2-8}
                                                                       & \multicolumn{1}{l|}{\multirow{4}{*}{PDMarks}} & P-only & 2,809 & -0.050 & 0.027 & 0.522 & $\mathrm{4.7e-38}$ \\ \cline{3-8}
                                                                       & \multicolumn{1}{l|}{}                         & C-only & -385 & -0.098 & 0.001 & 0.383 & $\mathrm{6.0e-8}$ \\ \cline{3-8}
                                                                       & \multicolumn{1}{l|}{}                         & R-only & 910 & 0.012 & -0.002 & -0.130 & $\mathrm{<1e-300}$ \\ \cline{3-8}
                                                                       & \multicolumn{1}{l|}{}                         & All-stage & -267 & -0.036 & 0.017 & 0.436 & \blue{$\mathbf{<1e-300}$} \\ \hline
\multirow{7}{*}{\begin{tabular}[c]{@{}l@{}}BP\\ NG45\end{tabular}} & \multicolumn{2}{l|}{Row-parity} & -731 & 0.001 & -0.006 & -0.006 & $\mathrm{1.2e-10}$ \\ \cline{2-8}
                                                                       & \multicolumn{2}{l|}{Buffer-insertion} & -568 & -0.013 & 0.000 & 0.000 & $\mathrm{4.5e-13}$ \\ \cline{2-8}
                                                                       & \multicolumn{2}{l|}{ICMarks} & 14 & -0.001 & 0.005 & 0.005 & $\mathrm{1.5e-11}$ \\ \cline{2-8}
                                                                       & \multicolumn{1}{l|}{\multirow{4}{*}{PDMarks}} & P-only & 136 & -0.003 & -0.005 & -0.005 & $\mathrm{1.8e-15}$ \\ \cline{3-8}
                                                                       & \multicolumn{1}{l|}{}                         & C-only & 25 & 0.009 & 0.002 & 0.002 & $\mathrm{6.0e-8}$ \\ \cline{3-8}
                                                                       & \multicolumn{1}{l|}{}                         & R-only & 187 & 0.001 & 0.003 & 0.003 & $\mathrm{<1e-300}$ \\ \cline{3-8}
                                                                       & \multicolumn{1}{l|}{}                         & All-stage & 137 & 0.006 & -0.002 & -0.002 & \blue{$\mathbf{<1e-300}$} \\ \hline
\multirow{6}{*}{\begin{tabular}[c]{@{}l@{}}JPEG\\ ASAP7\end{tabular}} & \multicolumn{2}{l|}{Row-parity} & 330 & 0.634 & 0.012 & 0.795 & $\mathrm{2.2e-19}$ \\ \cline{2-8}
                                                                       & \multicolumn{2}{l|}{Buffer-insertion} & 410 & 0.546 & 0.010 & 0.768 & $\mathrm{3.2e-24}$ \\ \cline{2-8}
                                                                       & \multicolumn{2}{l|}{ICMarks} & 854 & 1.250 & 0.009 & 0.773 & $\mathrm{1.4e-16}$ \\ \cline{2-8}
                                                                       & \multicolumn{1}{l|}{\multirow{3}{*}{PDMarks}} & P-only & 494 & 0.471 & 0.001 & 0.041 & $\mathrm{1.0e-25}$ \\ \cline{3-8}
                                                                       & \multicolumn{1}{l|}{}                         & C-only & 16 & 0.000 & 0.001 & 0.099 & $\mathrm{6.0e-8}$ \\ \cline{3-8}
                                                                       & \multicolumn{1}{l|}{}                         & All-stage & 419 & 0.431 & 0.005 & -0.053 & \blue{$\mathbf{6.2e-33}$} \\ \hline
\multirow{6}{*}{\begin{tabular}[c]{@{}l@{}}SweRV\\ ASAP7\end{tabular}} & \multicolumn{2}{l|}{Row-parity} & 51 & 0.048 & 0.022 & 1.832 & $\mathrm{3.6e-15}$ \\ \cline{2-8}
                                                                       & \multicolumn{2}{l|}{Buffer-insertion} & 130 & 0.047 & 0.028 & 1.993 & $\mathrm{5.3e-23}$ \\ \cline{2-8}
                                                                       & \multicolumn{2}{l|}{ICMarks} & 80 & 0.048 & 0.026 & 1.810 & $\mathrm{7.3e-12}$ \\ \cline{2-8}
                                                                       & \multicolumn{1}{l|}{\multirow{3}{*}{PDMarks}} & P-only & 9 & -0.015 & -0.004 & -0.405 & $\mathrm{1.0e-28}$ \\ \cline{3-8}
                                                                       & \multicolumn{1}{l|}{}                         & C-only & 126 & -0.023 & 0.012 & 1.491 & $\mathrm{6.0e-8}$ \\ \cline{3-8}
                                                                       & \multicolumn{1}{l|}{}                         & All-stage & 45 & -0.029 & -0.002 & -0.038 & \blue{$\mathbf{6.0e-36}$} \\ \hline
\multirow{6}{*}{\begin{tabular}[c]{@{}l@{}}CVA6\\ ASAP7\end{tabular}} & \multicolumn{2}{l|}{Row-parity} & 411 & 0.154 & -0.001 & -0.062 & $\mathrm{5.8e-11}$ \\ \cline{2-8}
                                                                       & \multicolumn{2}{l|}{Buffer-insertion} & 302 & 0.143 & -0.035 & -1.021 & $\mathrm{6.0e-15}$ \\ \cline{2-8}
                                                                       & \multicolumn{2}{l|}{ICMarks} & 389 & 0.155 & -0.023 & -0.751 & $\mathrm{7.6e-06}$ \\ \cline{2-8}
                                                                       & \multicolumn{1}{l|}{\multirow{3}{*}{PDMarks}} & P-only & 165 & 0.025 & 0.004 & 0.192 & $\mathrm{6.7e-52}$ \\ \cline{3-8}
                                                                       & \multicolumn{1}{l|}{}                         & C-only & 40 & -0.003 & -0.002 & 0.037 & $\mathrm{6.0e-8}$ \\ \cline{3-8}
                                                                       & \multicolumn{1}{l|}{}                         & All-stage & 243 & 0.036 & 0.000 & 0.058 & \blue{$\mathbf{1.4e-61}$} \\ \hline
\multirow{6}{*}{\begin{tabular}[c]{@{}l@{}}Ariane\\ ASAP7\end{tabular}} & \multicolumn{2}{l|}{Row-parity} & 121 & 0.164 & 0.008 & 0.000 & $\mathrm{1.2e-10}$ \\ \cline{2-8}
                                                                       & \multicolumn{2}{l|}{Buffer-insertion} & 78 & 0.175 & 0.010 & 0.000 & $\mathrm{1.4e-17}$ \\ \cline{2-8}
                                                                       & \multicolumn{2}{l|}{ICMarks} & 145 & 0.173 & 0.003 & 0.000 & $\mathrm{3.6e-15}$ \\ \cline{2-8}
                                                                       & \multicolumn{1}{l|}{\multirow{3}{*}{PDMarks}} & P-only & 60 & 0.007 & 0.004 & 0.000 & $\mathrm{3.7e-40}$ \\ \cline{3-8}
                                                                       & \multicolumn{1}{l|}{}                         & C-only & 172 & 0.000 & 0.003 & 0.000 & $\mathrm{6.0e-8}$ \\ \cline{3-8}
                                                                       & \multicolumn{1}{l|}{}                         & All-stage & 65 & 0.003 & 0.001 & 0.000 & \blue{$\mathbf{2.9e-45}$} \\ \hline
                                                                       
\end{tabular}}
\end{table}

Table~\ref{tab:ppa} reports the PPA changes and coincidence
probability for all eight designs. On NanGate45, the all-stage
\emph{PDMarks} flow changes routed wirelength by at most 5.5\,k$\mu$m and
total power by at most 5.9\,mW. Both changes are less than 1\% of the
corresponding reference values. The timing changes are also small,
with $|\Delta\mathrm{WNS}| \leq 0.03$\,ns and
$|\Delta\mathrm{TNS}| \leq 7.2$\,ns. On ASAP7, the all-stage flow
keeps $|\Delta\mathrm{WNS}| \leq 0.005$\,ns and
$|\Delta\mathrm{TNS}| \leq 0.06$\,ns. The placement-only, CTS-only,
and routing-only results further show that no individual carrier
dominates the implementation overhead.
Figure~\ref{fig:ppa} summarizes the PPA changes of \emph{PDMarks}
and the three baselines. Each gray point gives the result for one
design, and each bar reports the mean across the evaluated designs.
\emph{PDMarks} has comparable or smaller changes in routed wirelength,
power, WNS, and TNS. The mean changes remain close to zero, and the
per-design values remain within the same range as or below those of
the prior methods. These results show that distributing the
watermark across multiple PD stages does not induce a larger PPA
overhead than a single-stage watermark.

In terms of the coincidence probability $P_c$,
the prior methods rely on a single watermark carrier and obtain
$P_c$ values between approximately $10^{-6}$ and $10^{-32}$ across
the evaluated designs. Placement-only \emph{PDMarks} already provides
smaller $P_c$ than the baselines on most designs because it verifies
many independently selected local ordering claims. CTS-only
watermarking carries fewer claims and therefore has a larger $P_c$,
while routing-only watermarking provides very strong statistical
evidence because it aggregates observations over thousands of
key-selected nets.
For the all-stage flow, the joint coincidence probability
$P_c$ is below $10^{-32}$
on every design and below $10^{-300}$ on 
three NanGate45 designs.
Thus, the observed evidence across all
available stages is substantially less 
likely to arise by coincidence
than the evidence produced by the prior 
methods, while the PPA changes remain comparable.

\subsection{Full-flow Survival and Forensic Verification}
\label{subsec:survival}

The ownership proof must remain visible after the watermarked objects pass
through downstream PD stages. Table~\ref{tab:survival} reports extraction
rates at post-placement, post-CTS, post-global-route (post-GRT), 
and post-detailed-route (post-DRT) checkpoints. 
Placement evidence is measured by $r_P$, CTS evidence by
$r_C$, and routing evidence by the wrong-way-routing statistic $(Z_R,p_R)$.
The combined statistic $r_{\mathrm{all}}$ is computed from the stage
results available in the corresponding technology.
For reporting, we summarize the agreement among the stages
available at each checkpoint using
$r_{\mathrm{all}}=
  \frac{1}{|\mathcal{A}|}
  \sum_{s\in\mathcal{A}}r_s.$
  \label{eq:combined-extraction}
The statistic $r_{\mathrm{all}}$ is an evidence summary and does not
enter the ownership decision in Eq.~\eqref{eq:ownership_rule}.

All these stages remain verifiable at the final detailed
routing checkpoint. On NanGate45, the smallest 
final placement extraction rate is 0.969,
CTS extraction remains at 1.000, and the routing test is highly significant
for every design. On ASAP7, the smallest 
final placement extraction rate is 0.976, while
CTS extraction remains at 1.000.


\begin{table}[t]
\caption{Full-flow watermark survival and forensic verification across physical-design checkpoints.}
\label{tab:survival}
\centering
\scriptsize
\resizebox{0.48\textwidth}{!}{%
\begin{tabular}{|l|l|c|c|l|l|}
\hline
Design                                                                  & Evidence      & Post-Placement & Post-CTS & Post-GRT & Post-DRT \\ \hline
\multirow{4}{*}{\begin{tabular}[c]{@{}l@{}}JPEG\\ NG45\end{tabular}}     & $r_P$ & 1.000 & 0.961 & 0.969 & 0.969 \\ \cline{2-6}
                                                                        & $r_C$ & -- & 1.000 & 1.000 & 1.000 \\ \cline{2-6}
                                                                        & ($Z_R$, $p_R$, $r_R$) & -- & -- & 0, 0.5, 0 & 29.869, 2.499e-196, 1 \\ \cline{2-6}
                                                                        & $r_{\mathrm{all}}$ & 1.000 & 0.980 & 0.984 & 0.990 \\ \hline
\multirow{4}{*}{\begin{tabular}[c]{@{}l@{}}SweRV\\ NG45\end{tabular}}    & $r_P$ & 1.000 & 1.000 & 0.989 & 0.989 \\ \cline{2-6}
                                                                        & $r_C$ & -- & 1.000 & 1.000 & 1.000 \\ \cline{2-6}
                                                                        & ($Z_R$, $p_R$, $r_R$) & -- & -- & 0, 0.5, 0 & 38.910, $<$1e-300, 1 \\ \cline{2-6}
                                                                        & $r_{\mathrm{all}}$ & 1.000 & 1.000 & 0.994 & 0.996 \\ \hline
\multirow{4}{*}{\begin{tabular}[c]{@{}l@{}}Ariane\\ NG45\end{tabular}}   & $r_P$ & 1.000 & 1.000 & 1.000 & 1.000 \\ \cline{2-6}
                                                                        & $r_C$ & -- & 1.000 & 1.000 & 1.000 \\ \cline{2-6}
                                                                        & ($Z_R$, $p_R$, $r_R$) & -- & -- & 0, 0.5, 0 & 70.695, $<$1e-300, 1 \\ \cline{2-6}
                                                                        & $r_{\mathrm{all}}$ & 1.000 & 1.000 & 1.000 & 1.000 \\ \hline
\multirow{4}{*}{\begin{tabular}[c]{@{}l@{}}BP\\ NG45\end{tabular}}       & $r_P$ & 1.000 & 1.000 & 1.000 & 1.000 \\ \cline{2-6}
                                                                        & $r_C$ & -- & 1.000 & 1.000 & 1.000 \\ \cline{2-6}
                                                                        & ($Z_R$, $p_R$, $r_R$) & -- & -- & 0, 0.5, 0 & 47.635, $<$1e-300, 1 \\ \cline{2-6}
                                                                        & $r_{\mathrm{all}}$ & 1.000 & 1.000 & 1.000 & 1.000 \\ \hline
\multirow{3}{*}{\begin{tabular}[c]{@{}l@{}}JPEG\\ ASAP7\end{tabular}}    & $r_P$ & 1.000 & 1.000 & 1.000 & 1.000 \\ \cline{2-6}
                                                                        & $r_C$ & -- & 1.000 & 1.000 & 1.000 \\ \cline{2-6}
                                                                        & $r_{\mathrm{all}}$ & 1.000 & 1.000 & 1.000 & 1.000 \\ \hline
\multirow{3}{*}{\begin{tabular}[c]{@{}l@{}}SweRV\\ ASAP7\end{tabular}}   & $r_P$ & 1.000 & 1.000 & 1.000 & 1.000 \\ \cline{2-6}
                                                                        & $r_C$ & -- & 1.000 & 1.000 & 1.000 \\ \cline{2-6}
                                                                        & $r_{\mathrm{all}}$ & 1.000 & 1.000 & 1.000 & 1.000 \\ \hline
\multirow{3}{*}{\begin{tabular}[c]{@{}l@{}}Ariane\\ ASAP7\end{tabular}}  & $r_P$ & 1.000 & 0.992 & 0.992 & 0.992 \\ \cline{2-6}
                                                                        & $r_C$ & -- & 1.000 & 1.000 & 1.000 \\ \cline{2-6}
                                                                        & $r_{\mathrm{all}}$ & 1.000 & 0.996 & 0.996 & 0.996 \\ \hline
\multirow{3}{*}{\begin{tabular}[c]{@{}l@{}}CVA6\\ ASAP7\end{tabular}}    & $r_P$ & 1.000 & 0.976 & 0.976 & 0.976 \\ \cline{2-6}
                                                                        & $r_C$ & -- & 1.000 & 1.000 & 1.000 \\ \cline{2-6}
                                                                        & $r_{\mathrm{all}}$ & 1.000 & 0.988 & 0.988 & 0.988 \\ \hline
\end{tabular}}
\vspace{-1.5em}
\end{table}

\subsection{Wrong-Key Analysis}
\label{subsec:wrong-key}

\begin{figure}
    \centering
    \includegraphics[width=0.48\textwidth]{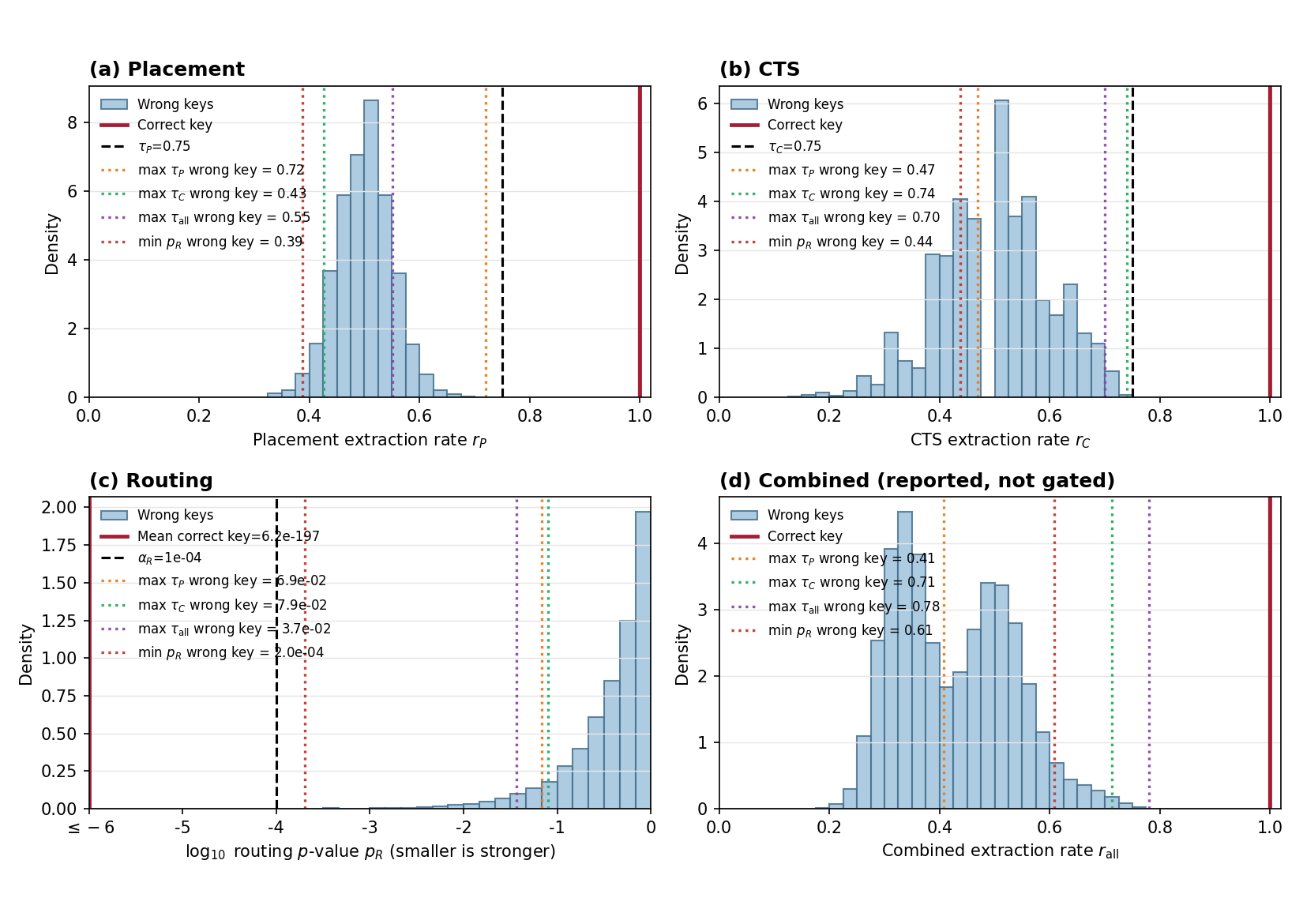}
  \caption{Wrong-key vs. correct-key verification.}
    \label{fig:wrong_key}
\vspace{-1em}
\end{figure}

This auxiliary experiment evaluates whether 
a claimant without the owner's key
can satisfy the ownership test.\footnote{This experiment
bypasses the commitment and certificate-authentication checks described in Section~\ref{subsec:ownership_verification}.
It evaluates only the stage-level statistics.} 
For each design, we verify the final watermarked layout once with the
correct key and with 5000 independently generated wrong keys.
In each wrong-key trial, the
placement and CTS objects and target values are reconstructed from the
public eligible sets using the trial key, while the routing-net
population is selected using the same public routing procedure.
Figure~\ref{fig:wrong_key} shows that the correct key produces
placement and CTS extraction rates near one, whereas wrong keys produce
rates near the random baseline of 0.5. For routing, 
the correct key yields
a small one-sided $p$-value\footnote{The mean value 
of $p_R$ across all designs using the correct
key is 6.2e-197.}, 
while the wrong-key $p$-values are
consistent with the null hypothesis.

Table~\ref{tab:wrong-key} summarizes the wrong-key results and the
resulting verification thresholds. Across all wrong keys, the
maximum placement and CTS extraction rates are 0.72 and 0.74,
respectively. We therefore use
\(\tau_P=\tau_C=0.75\). The minimum wrong-key routing \(p\)-value is
\(2\times10^{-4}\), and we use \(\alpha_R=10^{-4}\). 
Therefore, all wrong keys yield $r_R=0$, whereas the correct
key yields $r_R=1$ for every NanGate45 design.
Under these
thresholds, none of the wrong-key trials satisfies
the stage-level two-of-three rule on any design. 
These trials provide an empirical
characterization of random wrong-key behavior rather than a guarantee
against unlimited post hoc key search. Such retrospective search is
excluded by the pre-release key commitment.

\begin{table}[t]
\caption{Wrong-key analysis.}
\label{tab:wrong-key}
\centering
\scriptsize
\resizebox{0.43\textwidth}{!}{
\begin{tabular}{|c|cc|c|c|}
\hline
\multirow{2}{*}{\begin{tabular}[c]{@{}c@{}}Evidence \end{tabular}} & \multicolumn{2}{c|}{Wrong-key}                                             & \multirow{2}{*}{\begin{tabular}[c]{@{}c@{}}Correct-key\end{tabular}} & \multirow{2}{*}{Threshold} \\ \cline{2-3}
                                                                            & \multicolumn{1}{c|}{mean}                      & max                       &                                                                               &                            \\ \hline
$r_P$                                                                       & \multicolumn{1}{c|}{0.50}                     & 0.72                   & 1.00                                                                          & $\tau_P=0.75$              \\ \hline
$r_C$                                                                       & \multicolumn{1}{c|}{0.50}                     & 0.74                     & 1.00                                                                          & $\tau_C=0.75$              \\ \hline
$p_R$                                                                       & \multicolumn{1}{c|}{0.50}                     &  0.0002 (min)                    & 6.2e-197 (mean)                                & $\alpha_R=0.0001$            \\ \hline
$r_R$ & \multicolumn{1}{c|}{0.00}    & 0 & 1.00 & pass if $r_R=1$ \\ \hline
$r_{\mathrm{all}}$                                                                     & \multicolumn{1}{c|}{0.47}                     &0.78                   &        1.00                                                                    & reported (no gate)          \\ \hline
\end{tabular}}
\end{table}

\subsection{Parameter Sensitivity}
\label{subsec:param_sensitivity}
We sweep the key watermark parameters to 
analyze the parameter sensitivity of \emph{PDMarks}.
Each parameter is swept individually, while the 
remaining parameters are fixed to their default values.
For placement, we sweep the tuple span
$D_{\mathrm{p}}\in\{0.5,1.0,2.0\}\,\mu\mathrm{m}$,
the HPWL guard $\theta_{\mathrm{HPWL}}\in\{50,100,200\}$ DBU, and the
timing guard $\delta_{\mathrm{guard}}\in\{0.01,0.02,0.05\}\,\mathrm{ns}$.
For CTS, we sweep the LCB pairing distance
$R_{\max}\in\{25,50,100\}\,\mu\mathrm{m}$. For routing, we sweep the
watermark-net fraction $f\in\{0.02,0.05,0.10\}$ and the wrong-way bias
$\lambda_{\mathrm{wm}}\in\{10,100,1000\}$. 
We report the sweep on SweRV (NG45)
in Table~\ref{tab:sensitivity}. 

The \emph{placement} span primarily controls capacity. At
$D_{\mathrm{p}}=0.5\,\mu$m, no eligible same-row tuple is found.
Increasing the span to $1.0\,\mu$m and $2.0\,\mu$m increases the
eligible population to 36 and 528, respectively. 
Changing $\theta_{\mathrm{HPWL}}$ has little effect on either capacity
or PPA. Relaxing $\delta_{\mathrm{guard}}$ changes the accepted
placement solution and its PPA, but the selected population remains
unchanged.
For \emph{CTS}, increasing $R_{\max}$ enlarges the eligible LCB-pair
population from 4743 to 61700. The number of accepted CTS claims remains
24 because the CTS embedder uses a fixed claim budget. The extraction
rate remains 1.0 for all three settings, which indicates that a larger
candidate pool does not weaken the embedded claims.
For \emph{routing}, increasing $f$ raises the selected-net
population from 2151 to 5378 and 10758. The corresponding routing
test statistic increases from \(Z_R=38.9\) to \(85.7\) and \(123.0\),
respectively, because more routed segments contribute to the
population-level test. In contrast,
increasing \(\lambda_{\mathrm{wm}}\) from 10 to 100 and 1000 provides
little additional benefit. The three settings produce nearly identical
routing results, with \(Z_R=38.9\) and \(p_R<10^{-300}\), indicating
that \(\lambda_{\mathrm{wm}}=10\) is already sufficient for this
design.

Across the sweep, all settings with nonzero placement capacity retain
full extraction. The largest absolute changes are 0.031\,ns in WNS,
10.65\,ns in TNS, 4229\,$\mu$m in routed wirelength, and 0.164\,mW
in power. The routed-wirelength change remains below 0.12\% of the
reference value. These results show that the default parameters
provide sufficient capacity and strong verification without requiring
fine tuning.

\begin{table}[t]
\caption{Parameter sensitivity on SweRV (NG45).
PPA values are absolute changes from the unwatermarked reference.
$\Delta$WNS and $\Delta$TNS are reported in ns,
$\Delta$rWL in $\mu$m, and $\Delta$Power in mW.
Default values are in bold font.}
\label{tab:sensitivity}
\centering
\footnotesize
\resizebox{0.48\textwidth}{!}{%
\begin{tabular}{|l|l|c|c|c|c|c|c|c|}
\hline
Stage & Param & Value & $|E_s|$ & $|WM_s|$  & $\Delta$WNS (ns) & $\Delta$TNS (ns) & $\Delta$rWL ($\mu$m) & $\Delta$Power (mW) \\ \hline
\multirow{9}{*}{\begin{tabular}[c]{@{}l@{}}Place\\ ($r_P$)\end{tabular}}
  & \multirow{3}{*}{$D_{\mathrm{p}}$} & 0.5 & 0 & 0  & - & - & - & - \\ \cline{3-9}
  &                                      & 1.0 & 36 & 33  & +0.002 & +9.65 & +3460 & +0.164 \\ \cline{3-9}
  &                                      & {\bf 2.0} & {\bf 528} & {\bf 89}  & {\bf -0.026} & {\bf +7.15} & {\bf +1146} & {\bf -0.161} \\ \cline{2-9}
  & \multirow{3}{*}{$\theta_{\mathrm{HPWL}}$} & 50 & 416 & 89 & -0.026 & +7.16 &  +1146 &  -0.161 \\ \cline{3-9}
  &                                      & {\bf 100} & {\bf 528} & {\bf 89}  & {\bf -0.026} & {\bf +7.16} & {\bf +1146} & {\bf -0.161} \\ \cline{3-9}
  &                                      & 200 & 585 & 89 &-0.026 & +7.16 &  +1146 &  -0.161  \\ \cline{2-9}
  & \multirow{3}{*}{$\delta_{\mathrm{guard}}$} & 0.01 & 528 & 89  & -0.012 & +8.22 & +769 & +0.027 \\ \cline{3-9}
  &                                      & {\bf 0.02} & {\bf 528} & {\bf 89}  & {\bf -0.026} & {\bf +7.15} & {\bf +1146} & {\bf -0.161}  \\ \cline{3-9}
  &                                      & 0.05 & 528 & 89  & -0.012 & +8.22 & +769 & +0.027 \\ \hline
\multirow{3}{*}{\begin{tabular}[c]{@{}l@{}}CTS\\ ($r_C$)\end{tabular}}
  & \multirow{3}{*}{$R_{\max}$} & 25 & 4,743 & 24  & -0.019 & +9.17 & +4229 & -0.027 \\ \cline{3-9}
  &                             & {\bf 50} & {\bf 17950} & {\bf 24}  & {\bf -0.026} & {\bf +7.15} & {\bf +1146} & {\bf -0.161}  \\ \cline{3-9}
  &                             & 100 & 61,700 & 24 & -0.014 & +10.65 & +969 & -0.164 \\ \hline
\multirow{6}{*}{\begin{tabular}[c]{@{}l@{}}Route\\ ($Z_R$)\end{tabular}}
  & \multirow{3}{*}{$f$} & {\bf 0.02} & {\bf 107577} & {\bf 2151} & {\bf -0.026} & {\bf +7.15} & {\bf +1146} & {\bf -0.161}  \\ \cline{3-9}
  &                      & 0.05 & 107577 & 5378  & -0.023 & -0.77 & +769 & +0.027 \\ \cline{3-9}
   &                    & 0.10 & 107577 & 10758  & -0.031 & -1.31 & +1538 & +0.027 \\ \cline{2-9}
  & \multirow{3}{*}{$\lambda_{\mathrm{wm}}$} & 10 & 107577 & 2151  & -0.026 & +7.15 & +1146 & -0.161  \\ \cline{3-9}
  &                      & {\bf 100} & {\bf 107577} & {\bf 2151} & {\bf -0.026} & {\bf +7.15} & {\bf +1146} & {\bf -0.161}  \\ \cline{3-9}
  &                      & 1000 & 107577 & 2151  & -0.026 & +7.15 & +1146 & -0.161  \\ \hline
\end{tabular}
}
\end{table}

\section{Attack Evaluation}
\label{sec:attack_evaluation}
This section evaluates the robustness of \emph{PDMarks} under white-box
attacks that attempt to remove or weaken the ownership
evidence. The attacker is assumed to know the watermarking algorithm,
source code, flow checkpoints, public eligibility rules,
and the verification procedure, but not the
secret key. Therefore, they cannot reconstruct the key-selected
watermark objects. 
The attacker may possess the public ciphertext $C_\Gamma$, but cannot
recover the certified placement and CTS claims without $K$.

We evaluate blind and targeted attacks on placement, CTS, and routing.
For each attack, we report the post-attack placement and CTS
extraction rates, the routing statistic, and the PPA changes relative
to the stolen watermarked design. An attack succeeds only when 
fewer than two available
stages pass their corresponding verification tests  
and the attacked
layout remains legal without excessive PPA degradation.

\subsection{Blind Attacks}
\label{subsec:blind_attack}

\begin{figure*}
    \centering
    \includegraphics[width=0.85\textwidth]{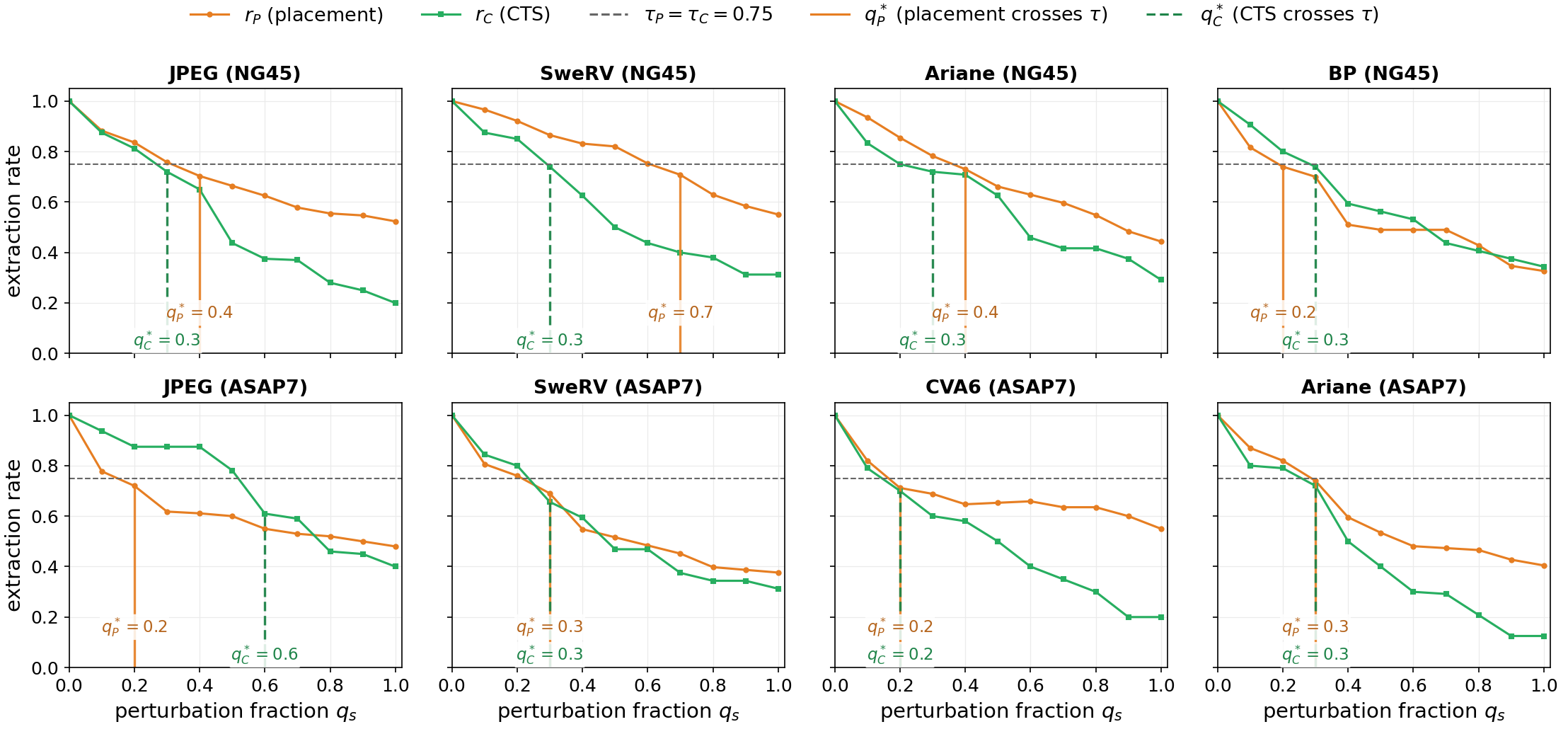}
  \caption{Placement and CTS extraction rates under blind perturbation.
The vertical lines indicate the smallest perturbation fraction at
which each extraction rate falls below its threshold.}
    \label{fig:blind_extraction}
\vspace{-1em}
\end{figure*}

\begin{figure}
    \centering
    \includegraphics[width=0.5\textwidth]{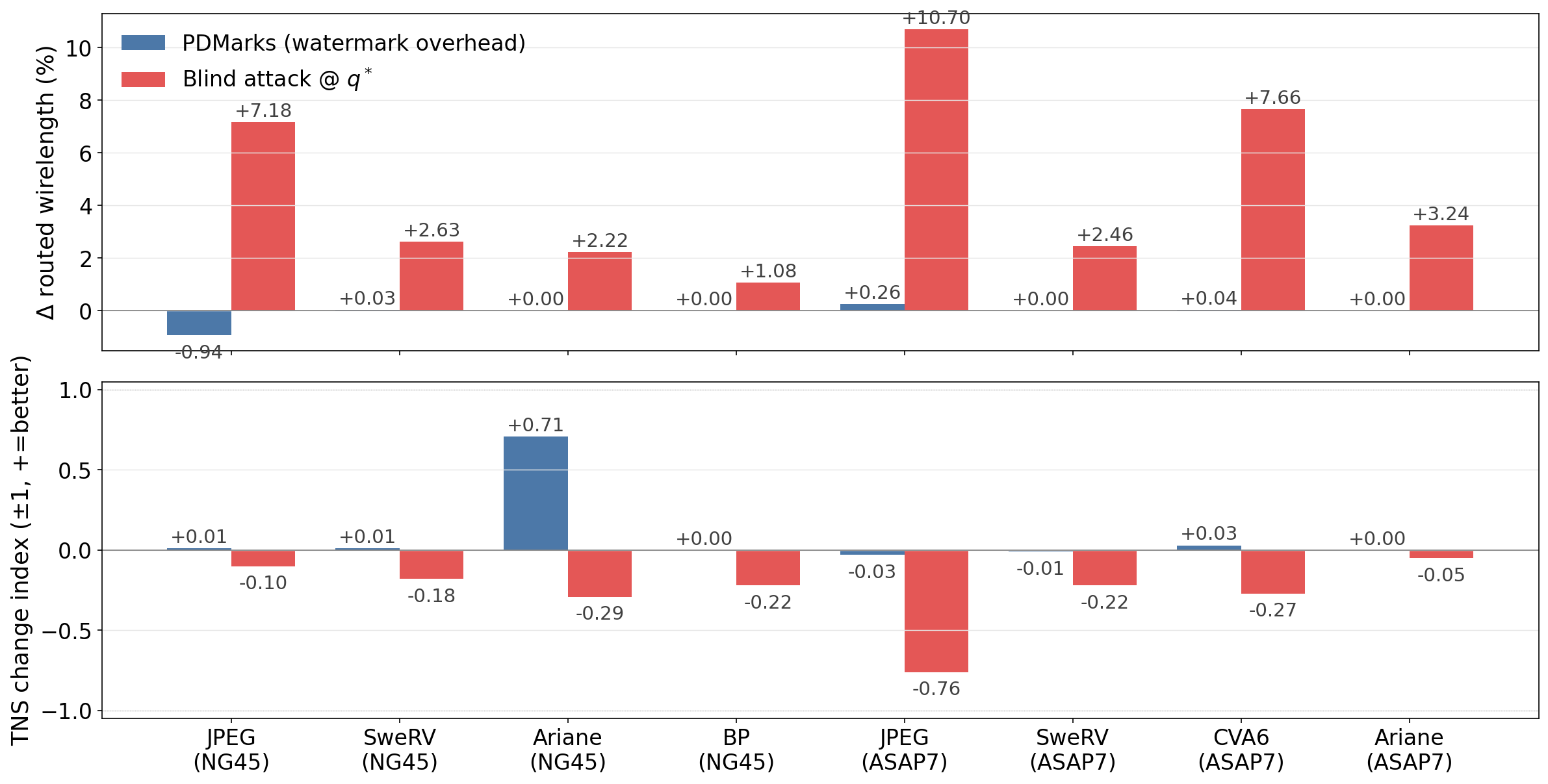}
  \caption{PPA cost of blind placement and CTS attacks at the minimum
perturbation fraction that causes a stage-level verification failure.
The blue bars show the original watermark overhead, and the red bars
show the additional attack cost.}
    \label{fig:blind_cost}
\vspace{-1em}
\end{figure}

A \emph{blind attack} randomly perturbs watermark-eligible physical 
objects without knowing whether they carry watermark claims.
For each stage $s\in\{P,C,R\}$, let ${E}_s$ denote the public
eligible object set and let $\mathrm{WM}_s\subseteq {E}_s$ denote
the unknown key-selected watermark set. The attacker perturbs a fraction
$q_s$ of ${E}_s$, where
$q_s \in [10\%, 100\%]$ in increments
of 10\%.
Since $\mathrm{WM}_s$ is selected pseudorandomly from ${E}_s$,
the expected fraction of watermark objects affected by a blind
candidate-level attack is also $q_s$. Thus, to remove a constant
fraction of the watermark, the attacker must perturb a comparable
fraction of the eligible design objects. We evaluate whether this level
of perturbation is compatible with retaining the stolen physical
implementation.

\textbf{Placement attack.}
The attacker reconstructs the public set of eligible
same-row cell tuples and then selects
a fraction $q_P$ of these tuples and changes their local ordering. For
two-cell tuples, the attacker swaps the two cells. For three-cell
tuples, the attacker applies a random non-identity permutation to the
three cells. After perturbation, the design is legalized with the same
maximum-displacement limit used in the owner flow, followed by timing
analysis and the remaining stages. 

\textbf{CTS attack.}
The attacker reconstructs the public set of compatible
leaf-clock-buffer (LCB) pairs. For a fraction $q_C$ of these pairs, the
attacker locally perturbs the clock-sink assignment by moving one nearby
sequential sink between compatible LCBs. 
After the perturbation, the attacker performs local clock repair and
incremental timing repair. 

Figure~\ref{fig:blind_extraction} reports placement and CTS extraction
as the perturbation fraction increases from 0.1 to 1.0. Both rates
decrease monotonically because a larger fraction of the eligible
population is modified. The vertical lines mark the smallest
perturbation fractions $q_P^*$ and $q_C^*$ that reduce the
corresponding extraction rate below 0.75. Across the designs,
$q_P^*$ ranges from 0.2 to 0.7, and $q_C^*$ ranges from 0.2 to
0.6. Thus, a blind attacker must perturb a substantial fraction of
the publicly eligible tuples or LCB pairs before the corresponding
stage-level claim fails.
Figure~\ref{fig:blind_cost} reports the PPA cost at the
first perturbation point that causes either placement or CTS
verification to fail. The
attack increases routed wirelength on every design, with changes from
approximately 1\% to more than 10\% at the failure point. The TNS
changes are also larger than the original watermark overhead on all
designs. Therefore, the attacker can suppress a stage-level claim only
after introducing changes that are much larger than those required to
embed the watermark.


\textbf{Routing attack.}
The routing attack differs from the placement and CTS attacks because
the routing watermark is detected from aggregate wrong-way wiring
statistics. A strong attacker may rank nets by their observed
wrong-way usage and identify nets that are likely to belong to the
watermarked population. To evaluate this worst-case scenario, we assume
that the attacker can identify the watermarked nets.
We do not count full rerouting as a successful attack,
because it discards the routed interconnect that constitutes part of the
protected PD IP. Instead, we consider a local preservation attack in
which the attacker rips up and reroutes only the identified
watermarked nets without the watermark wrong-way penalty, followed by
DRC cleanup. 

\begin{figure}
    \centering
    \includegraphics[width=0.4\textwidth]{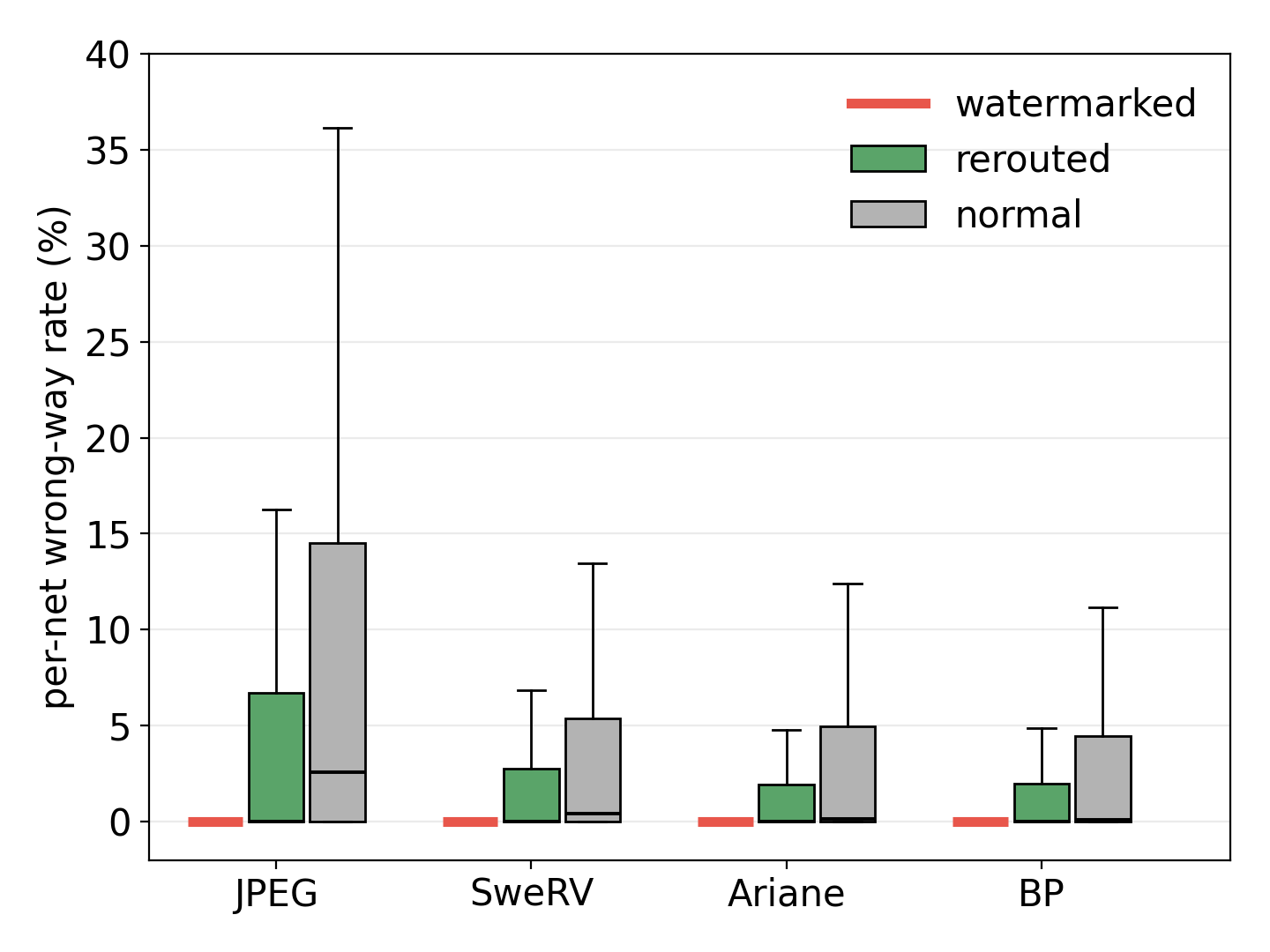}
  \caption{Wrong-way routing fractions of watermarked, locally rerouted,
and unselected nets under the routing removal attack. }
    \label{fig:routing_attack}
\vspace{-1.5em}
\end{figure}

Figure~\ref{fig:routing_attack} compares the wrong-way fractions of
the watermarked nets, the same nets after local rip-up and reroute, and
the unselected eligible nets. Watermark embedding reduces wrong-way
routing on the selected population to nearly zero. 
After local rerouting without the watermark bias, 
the rerouted nets move toward the 
wrong-way distribution of ordinary nets, but remain statistically 
distinguishable from the unselected population.
The $p_R$ values are $6.8\times10^{-10}$, $1.5\times10^{-5}$,
$3.1\times10^{-20}$, and $1.8\times10^{-9}$ for 
JPEG, SweRV, Ariane, and BP, respectively.
Accordingly, $r_R=1$ for all four designs, demonstrating that the
routing watermark survives this attack.
This resilience arises because local rerouting preserves 
the surrounding routed design, 
which restricts the available tracks and layer choices for 
the rerouted nets and prevents them from reaching the distribution 
produced by unconstrained routing.

\subsection{Targeted Attacks}
\label{subsec:targeted_attack}

A \emph{targeted attack} uses observable design features 
to rank watermark-eligible objects according to their 
likelihood of carrying watermark evidence, and then 
perturbs the highest-ranked objects.
This attack is more effective than the 
blind attack in Section~\ref{subsec:blind_attack} only when the
ranking distinguishes selected watermark objects from unselected
eligible objects.
We evaluate a strong oracle-assisted version of this attack.
For each stage, we reconstruct the eligible set $E_s$ 
and train a five-fold cross-validated RandomForest classifier.
Objects in the watermark set ${WM}_s$ are labeled positive and
the remaining eligible objects are labeled negative. 
A practical keyless attacker does not have access 
to these labels because membership in
$WM_s$ depends on the secret key. The resulting classifier therefore
gives the attacker more information than permitted by our threat
model and provides a conservative evaluation of feature-based
targeting.

For \emph{placement}, the feature vector includes local
placement density, row index and normalized position, cell widths and
heights, cell area, and fanout statistics of the tuple members.
For \emph{CTS}, the feature vector includes LCB fanout,
sequential-sink count, non-sequential fanout count, clock-tree depth,
estimated clock latency, local skew, load capacitance, and number of
nearby boundary flip-flops. 
For \emph{routing}, the features include net degree, bounding-box dimensions,
routed wirelength, via count, wrong-way segment count, wrong-way
fraction, and layer-usage distribution.

Table~\ref{tab:targeted_auc} reports the cross-validated AUC 
together with the precision and recall at top-$K$, 
where $K=|\mathrm{WM}_s|$. The
random-selection rate is $|\mathrm{WM}_s|/|E_s|$. 
An AUC close to $0.5$, 
together with precision at top-$K$ close to
$|WM_s|/|E_s|$, indicates that the ranking provides little advantage
over random selection.
Therefore, the targeted attack
reduces to the blind attack and we do not separately tabulate its
extraction.

For placement and CTS, the diagnostics in Table~\ref{tab:targeted_auc}
fall close to the random baseline on every design. The mean AUC is
$0.54$ for placement and $0.55$ for CTS; the mean top-$K$ precision is
$0.18$ and $0.01$, respectively; and the mean top-$K$ recall is the same
as the precision since $K\!=\!|\mathrm{WM}_s|$. 
The ranking therefore selects watermarked objects at a rate close to a
random subset, and the
resulting targeted attacks behave similarly to the corresponding
blind attacks in Figure~\ref{fig:blind_cost}.
For routing, the wrong-way features produce a mean
AUC of 0.77, so the ranking enriches the selected nets but does not
identify them exactly. The stronger oracle attack in
Figure~\ref{fig:routing_attack}, which is given the exact watermarked nets,
still does not reproduce the distribution obtained by routing
normal nets without watermarks.


\begin{table}[t]
\caption{Classifier diagnostics for targeted attacks.
AUC, precision at top-$K$, and recall at top-$K$ are obtained by
five-fold cross-validation over the eligible set $E_s$, where
$K=|WM_s|$. Routing is reported on NanGate45 only.}
\label{tab:targeted_auc}
\centering
\scriptsize
\resizebox{0.42\textwidth}{!}{%
\begin{tabular}{|c|l|c|c|c|c|}
\hline
Stage & Design & random & AUC & prec@$K$ & rec@$K$ \\ \hline
\multirow{9}{*}{Placement}
& JPEG (NG45)    & 0.182 & 0.476 & 0.172 & 0.172 \\
& SweRV (NG45)   & 0.169 & 0.528 & 0.180 & 0.180 \\
& Ariane (NG45)  & 0.140 & 0.577 & 0.169 & 0.169 \\
& BP (NG45)      & 0.195 & 0.586 & 0.200 & 0.200 \\
& JPEG (ASAP7)   & 0.168 & 0.514 & 0.172 & 0.172 \\
& SweRV (ASAP7)  & 0.153 & 0.502 & 0.161 & 0.161 \\
& Ariane (ASAP7) & 0.159 & 0.592 & 0.191 & 0.191 \\
& CVA6 (ASAP7)   & 0.148 & 0.524 & 0.153 & 0.153 \\\cline{2-6}
& mean            &       & 0.537 & 0.175 & 0.175 \\ \hline
\multirow{9}{*}{CTS}
& JPEG (NG45)    & 0.003 & 0.617 & 0.005 & 0.005 \\
& SweRV (NG45)   & 0.001 & 0.628 & 0.063 & 0.063 \\
& Ariane (NG45)  & 0.001 & 0.521 & 0.000 & 0.000 \\
& BP (NG45)      & 0.001 & 0.545 & 0.000 & 0.000 \\
& JPEG (ASAP7)   & 0.016 & 0.556 & 0.000 & 0.000 \\
& SweRV (ASAP7)  & 0.009 & 0.668 & 0.031 & 0.031 \\
& CVA6 (ASAP7)   & 0.016 & 0.409 & 0.000 & 0.000 \\
& Ariane (ASAP7) & 0.006 & 0.491 & 0.000 & 0.000 \\\cline{2-6}
& mean            &       & 0.554 & 0.013 & 0.013 \\ \hline
\multirow{5}{*}{Routing}
& JPEG (NG45)    & 0.020 & 0.773 & 0.106 & 0.106 \\
& SweRV (NG45)   & 0.020 & 0.774 & 0.140 & 0.140 \\
& Ariane (NG45)  & 0.020 & 0.805 & 0.181 & 0.181 \\
& BP (NG45)      & 0.020 & 0.763 & 0.090 & 0.090 \\\cline{2-6}
& mean            &       & 0.771 & 0.129 & 0.129 \\ \hline
\end{tabular}}
\end{table}

\subsection{Comparison with Baseline Watermarks}
\label{subsec:baseline_attack}

\begin{table}[t]
\caption{Blind and targeted attack results on baselines.}
\label{tab:baseline_attack}
\centering
\scriptsize
\resizebox{0.4\textwidth}{!}{%
\begin{tabular}{|l|c|c|c|c|c|}
\hline
\multirow{2}{*}{Method} &
\multirow{2}{*}{AUC} & \multicolumn{2}{c|}{$q_s\!=\!0.1$} &
\multicolumn{2}{c|}{$q_s\!=\!0.5$} \\ \cline{3-6}
  & & $r_B$ & $r_T$ & $r_B$ & $r_T$ \\ \hline
\emph{PDMarks} (place)  & 0.54 & 0.85 & 0.77 & 0.62 & 0.58 \\ \hline
Row-parity             & 0.49 & 0.97 & 0.95 & 0.72 & 0.66 \\ \hline
Buffer-insertion     & 0.64 & 0.90 & 0.67 & 0.54 & 0.37 \\ \hline
ICMarks              & 0.99 & 0.93 & 0.03 & 0.58 & 0.02 \\ \hline
\end{tabular}}
\end{table}

We next apply the same
blind and targeted attacks to the baseline watermarking
approaches. For each baseline, we construct the corresponding 
eligible set, train the same cross-validated classifier
on observable features, and re-verify the post-attack 
extraction with the method's own verifier. 
The feature sets used for the targeted attacks are provided
in our open-source GitHub repository~\cite{PDMarks_repo}.

Table~\ref{tab:baseline_attack} reports classifier
AUC and extraction rate under the blind ($r_B$) 
and targeted ($r_T$) attacks. All these values
are averaged across the designs.
Because carrier types and eligible
populations differ across methods, 
the table compares attack behavior at
matched perturbation fractions rather 
than matched numbers of modified
physical objects. 
For the \emph{parity-based} methods, such as \emph{PDMarks} placement
and Row-parity, the watermarked cells are statistically indistinguishable
from the eligible population ($AUC \approx 0.5$), so the targeted attacker does
no better than the blind one ($r_T\!\approx\!r_B$). In contrast, ICMarks 
places every marked cell inside a single region $R_w$ that is directly visible in
the layout. The classifier identifies it almost perfectly ($AUC=0.99$),
and the targeted attacker erases essentially the entire watermark at
$q^\ast\!=\!0.1$ ($r_T\!\le\!0.03$). Buffer-insertion exposes an
intermediate footprint (AUC$=0.64$).
In conclusion, 
\emph{PDMarks} provides the strongest overall
tradeoff among the evaluated methods. It achieves low implementation
overhead, stronger ownership evidence across multiple
PD stages, and greater
resilience to both blind and targeted attacks.

\section{Conclusion}
\label{sec:conclusion}

We have presented \emph{PDMarks}, a Kerckhoffs-compliant 
watermarking framework for physical design IP protection.  
\emph{PDMarks} embeds keyed ownership evidence across 
placement, CTS, and routing by using ordinary degrees 
of freedom in physical implementation.
The framework is integrated into OpenROAD-flow-scripts 
and supports read-only forensic verification from a 
suspect layout. Experiments across NanGate45 and ASAP7 
benchmarks evaluate watermark capacity, PPA overhead, 
full-flow survival, wrong-key false positives, and resilience 
against blind and targeted attacks. The results show that 
\emph{PDMarks} provides stronger ownership 
evidence than prior watermarking approaches while preserving 
implementation quality.
Future work includes extending the framework to additional 
physical artifacts, such as power-grid synthesis 
and timing ECOs, and developing stronger statistical 
combination rules for heterogeneous ownership evidence
across PD stages.

\section{Generative AI Acknowledgment}

ChatGPT5.6 was used in a proofreading pass and to tighten the treatment of key commitment 
in Section IV.D. No other use was made of generative AI.


\begin{IEEEbiography}[{\includegraphics[width=1in,height=1.25in,clip,keepaspectratio]{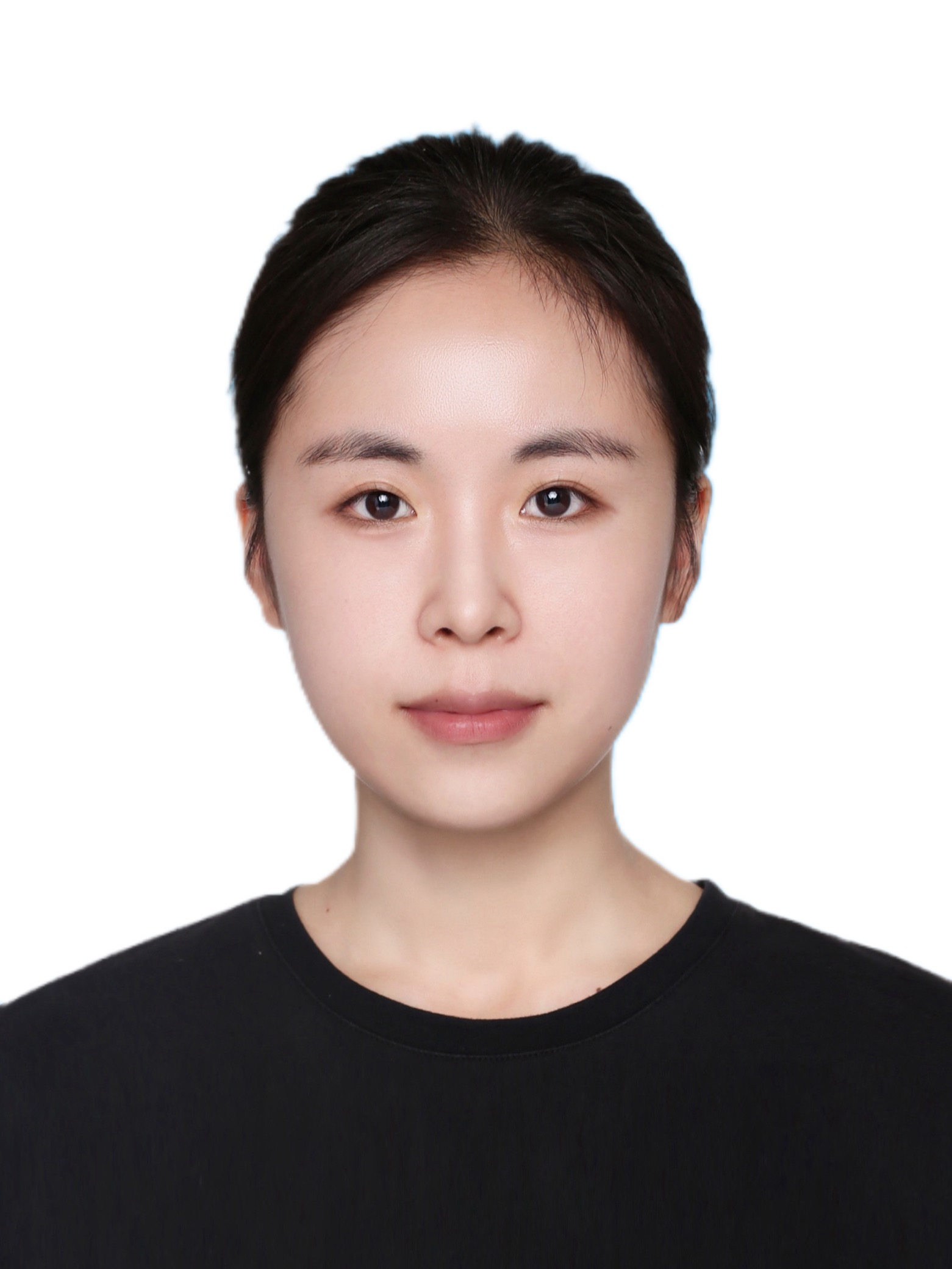}}]
    {Yiting Liu} is a postdoctoral scholar at the University of California San Diego, CA, USA. She received her Ph.D. degree in Computer Science from Fudan University, Shanghai, China, in 2024. Her current research interests include VLSI physical design and AI/ML for EDA and IC design.
\end{IEEEbiography}
\vspace*{-2\baselineskip}

\begin{IEEEbiography}[{\includegraphics[width=1in,height=1.25in,clip,keepaspectratio]{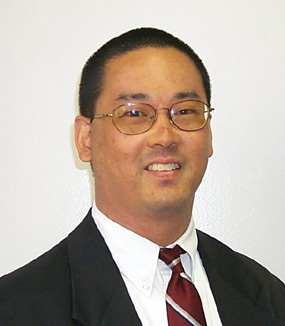}}]
    {Andrew B. Kahng} is Distinguished Professor of CSE and ECE
    at the University of California, San Diego. His interests include 
    IC physical design, the design-manufacturing interface,  large-scale combinatorial optimization, and AI/ML for EDA and IC design. He received the Ph.D. degree in Computer Science from the University of California, San Diego.
\end{IEEEbiography}

\end{document}